\documentclass[twocolumn]{aastex62}
\usepackage{graphicx,url}
\usepackage{amsmath,amssymb}
\usepackage{hyperref}
\hypersetup{colorlinks,
	linkcolor=blue, 
	urlcolor=magenta, 
	anchorcolor=blue,
	citecolor=blue
}
\usepackage{bm}
\usepackage{xcolor}
\usepackage{mathrsfs}
\usepackage{natbib}

\begin{document}

\title{\large\bfseries Reconstructing Functions and Estimating Parameters with Artificial Neural \\ Networks: A Test with the Hubble Parameter and SNe Ia}

\correspondingauthor{Jun-Qing Xia}
\email{xiajq@bnu.edu.cn}

\author{Guo-Jian Wang}
\affil{Department of Astronomy, Beijing Normal University, Beijing 100875, China}

\author{Xiao-Jiao Ma}
\affil{Department of Astronomy, Beijing Normal University, Beijing 100875, China}

\author{Si-Yao Li}
\affil{College of Information Science and Technology, Beijing Normal University, Beijing 100875, China}

\author{Jun-Qing Xia}
\affil{Department of Astronomy, Beijing Normal University, Beijing 100875, China}

\begin{abstract}
In this work, we propose a new nonparametric approach for reconstructing a function from observational data using an Artificial Neural Network (ANN), which has no assumptions about the data and is a completely data-driven approach. We test the ANN method by reconstructing functions of the Hubble parameter measurements $H(z)$ and the distance-redshift relation $D_L(z)$ of Type Ia supernovae. We find that both $H(z)$ and $D_L(z)$ can be reconstructed with high accuracy. Furthermore, we estimate cosmological parameters using the reconstructed functions of $H(z)$ and $D_L(z)$ and find the results are consistent with those obtained using the observational data directly. Therefore, we propose that the function reconstructed by ANN can represent the actual distribution of observational data and can be used for parameter estimation in further cosmological research. In addition, we present a new strategy for training and evaluating the neural network, and a code for reconstructing functions using ANN has been developed and is available for \href{https://github.com/Guo-Jian-Wang/refann}{{\it download}}.
\end{abstract}

\section{Introduction}

The accelerating expansion of the universe is a major discovery in modern cosmology. Many dynamic mechanisms have been proposed to explain this phenomenon, such as dark energy, modified gravity, and violation of the cosmological principle. However, the nature of this phenomenon is still unknown. The expansion of the universe can be quantitatively studied through various cosmological observations. It is an important issue to obtain information on the universe directly from the observational data without introducing any hypotheses (such as a cosmic model), which is also very important for understanding the nature of cosmic evolution and the theory of gravity. However, the dependence of the result obtained from observations on cosmological models is a thorny problem in cosmological research.

A Gaussian Process (GP) is a fully Bayesian approach that describes a distribution over functions and is a generalization of Gaussian distributions to function space \citep{Seikel:2012a}. It is a powerful nonlinear interpolating tool without assuming a model or parameterization and is widely used in cosmology literature, such as the construction of the dark energy equation of state \citep{Seikel:2012a,Seikel:2012b,Seikel:2013,Yahya:2014,Yang:2015,Wang:2019}, the reconstruction of cosmic expansion \citep{Montiel:2014,Li:2016a,Zhang:2016,WangDeng:2017b}, the test of cosmic curvature \citep{Cai:2016,Li:2016b,Yu:2016,Rana:2017,Wei:2017,Yu:2018,Wang:2019}, the estimation of the Hubble constant \citep{Busti:2014,Gomez-Valent:2018}, the tests of cosmic growth and matter perturbations \citep{Shafieloo:2013,Gonzalez:2017}, and the test of the distance duality relation \citep{ZhangYi:2014,Santos-da-Costa:2015,Li:2018,Melia:2018,Yang:2019}. In these papers, functions of the Hubble parameter with respect to the redshift and the distance-redshift relation are frequently reconstructed from expansion rate measurements and Type Ia supernovae (SNe Ia), respectively. Moreover, the derivatives and integrals of these functions are obtained for other applications, such as studying the evolution of dark energy and the constraint on the cosmic curvature.

However, \citet{Zhou:2019} recently proposed that the GP should be used with caution for the reconstruction of the Hubble parameter and SNe Ia. Moreover, \citet{Wei:2017} and \citet{Wang:2017} also found that GP is sensitive to the fiducial Hubble constant $H_0$, and the results are greatly influenced by $H_0$, which may imply the unreliability of the GP in the reconstruction of $H(z)$. In the analysis of GP, the errors in the observational data are assumed to obey a Gaussian distribution \citep{Seikel:2012a}. However, the actual observations might not obey Gaussian distributions. Thus, this may be a strong assumption for reconstructing functions from data.

An artificial neural network (ANN) is a machine learning method and has been proven to be a ``universal approximator'' that can represent a great variety of functions \citep{Cybenko:1989,Hornik:1991}. This powerful property of neural networks makes it widely used in regression and estimation tasks. With the development of computer hardware in the last decade, ANN is now capable of containing deep layers and training with a large amount of data. Recently, methods based on ANNs have shown outstanding performance in solving cosmological problems in both accuracy and efficiency. For example, it performs excellently in analyzing gravitational waves (GWs; \citet{LiXiangru:2017,George:2018}), estimating parameters of the 21 cm signal \citep{Shimabukuro:2017,Schmit:2018}, discriminating cosmological and reionization models \citep{Schmelzle:2017,Hassan:2018}, estimating cosmological parameters \citep{Fluri:2018,Fluri:2019,Ntampaka:2019,Ribli:2019}, searching and estimating parameters of strong gravitational lenses \citep{Jacobs:2017,Petrillo:2017,Hezaveh:2017,Pourrahmani:2018,Schaefer:2018}, classifying the large-scale structure of universe \citep{Aragon-Calvo:2019}, researching the background evolution of the universe \citep{WangDeng:2017a,Arjona:2019}, and studying the evolution of dark energy models \citep{Escamilla-Rivera:2019}.

An ANN is a collection of processing units designed to identify underlying relationships in input data, which is a completely data-driven method; hence, there are no assumptions of Gaussian distribution for the data. Therefore, the model established by ANN can describe the distribution of the input data correctly if an appropriate network is selected. In this work, based on ANN, we propose a new nonparametric method to reconstruct functions from data. We test this method by reconstructing functions of the Hubble parameter $H(z)$ and the distance-redshift relation $D_L(z)$ of SNe Ia. 

This paper is organized as follows: in section \ref{sec:methodology}, we take the Hubble parameter as an example to illustrate the process of the ANN method that is used to reconstruct functions from data. We first introduce the ANN method used in this work, then the method of simulating the Hubble parameter, and finally the process of reconstructing functions of $H(z)$ with the ANN method. In section \ref{sec:reconstruct_Hz}, we reconstruct functions of $H(z)$ from the observational data with the ANN method. Section \ref{sec:reconstruct_DL} presents the application of the ANN method in the reconstruction of the distance-redshift relation of SNe Ia. In section \ref{sec:compare_other_NN}, we compare the ANN method with other neural networks. In section \ref{sec:discussion}, a discussion about the ANN method is presented. Finally, a conclusion is shown in section \ref{sec:conclusion}.

\section{Methodology}\label{sec:methodology}

In this section, we first introduce the ANN method that is used in this work and then takes the Hubble parameter as an example to illustrate the process of reconstructing a function from data. Based on PyTorch,\footnote{\url{https://pytorch.org/docs/master/index.html}} an open source optimized tensor library for deep learning, we have developed a code for reconstructing functions from data called Reconstruct Functions with ANN (ReFANN). It can be used to reconstruct a function from a given data set using CPUs or GPUs.

\subsection{Artificial Neural Network}

\begin{figure}
	\centering
	\includegraphics[width=0.48\textwidth]{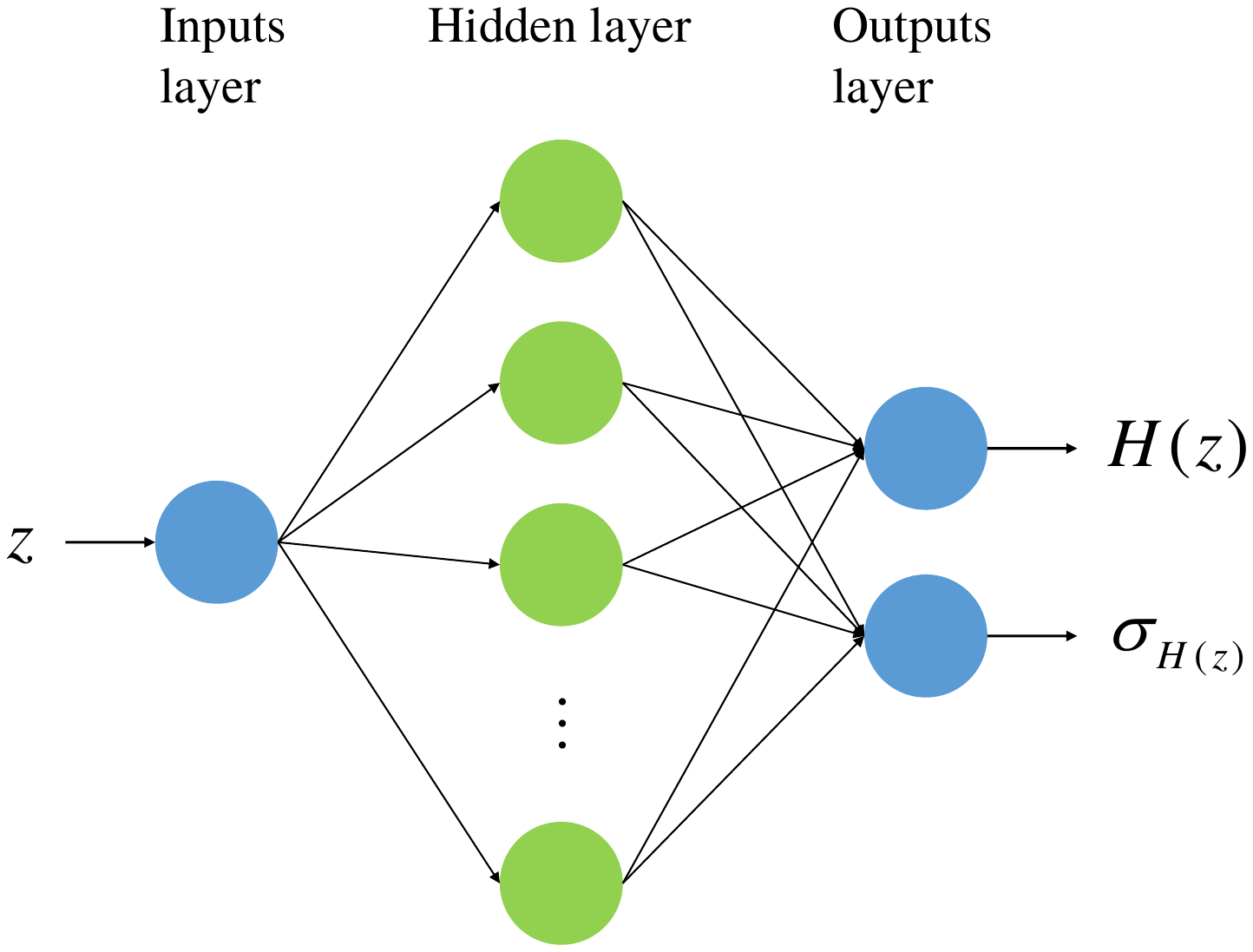}
	\caption{The general structure of the ANN used in this work. The input is the redshift $z$ of a Hubble parameter $H(z)$, and the outputs are the corresponding value and error of $H(z)$.}\label{fig:nn_model}
\end{figure}

An ANN, also called a Neural Network (NN), is a mathematical model that is inspired by the structure and functions of biological neural networks. The main purpose of an ANN is to construct an approximate function that associates input data with output data. An ANN generally consists of an input layer, hidden layers, and an output layer. The general structure of an ANN with one hidden layer used in this work is shown in Figure \ref{fig:nn_model}. Each layer accepts a vector, the elements of which are called neurons, from the previous layer as input, then apply a linear transformation and a nonlinear activation on the input, and finally propagates the current result to the next layer. Formally, in a vectorized style,
\begin{equation}\label{eq:def1}
\bm z_{i+1} = \bm x_{i}W_{i+1} + \bm b_{i+1},
\end{equation}
\begin{equation}\label{eq:def2}
\bm x_{i+1} = f(\bm z_{i+1}),
\end{equation}
where $\bm{x}_{i}$ is the input row vector of the $i$th layer, $W_{i+1}$ and $\bm b_{i+1}$ are linear weights and biases to be learned, $\bm z_{i+1}$ is the intermediate vector after linear transformation, and $f$ the elementwise nonlinear function. The output layer only takes linear transformations. In real implementation, for $n$ inputs of $x$ with shape $1\times n$ and $m$ neurons, the matrix $W$ has shape $n\times m$ and $b$ has shape $1\times m$. Thus, $z$ has the shape $1\times m$. In this work, we take the Exponential Linear Unit (ELU; \citet{Clevert:2015}) as the activation function, which has the form
\begin{align}\label{eq:elu}
&f(x) = 
\begin{cases}
x & x > 0 \\
\alpha (\exp(x)-1) & x \leq 0
\end{cases} \quad ,
\end{align}
where $\alpha$ is the hyperparameter that controls the value to which an ELU saturates for negative net inputs. In our network model, $\alpha$ is set to 1.

NNs are usually designed to process a batch of data simultaneously. Consider a matrix $X\in \mathbb{R}^{m\times n}$ where $m$ is the batch size and each row of $X$ is an independent input vector, then Equations (\ref{eq:def1}) and (\ref{eq:def2}) are replaced by the following batch-processed version:
\begin{equation}\label{eq:matrixdef1}
Z_{i + 1} = X_{i}W_{i+1}+B_{i + 1},
\end{equation}
\begin{equation}\label{eq:matrixdef2}
X_{i+1} = f(Z_{i+1}),
\end{equation}
where $B_{i+1}$ is the vertically replicated matrix of $\bm b_{i+1}$ in Equation \eqref{eq:def1}. An NN equals a function $f_{W, b}$ on input $X$. In supervised learning tasks, every input data is labeled corresponding to a ground-truth target $Y \in \mathbb{R}^{m\times p}$. The training process of a network is to minimize the difference between the predicted result $\hat{Y} = f_{W, b}(X)$ and the ground truth, which is quantitatively mapped with a loss function $\mathcal{L}$, by optimizing the parameters $W$ and $\bm b$. The least absolute deviation is used as the loss function in this work and has the following form:
\begin{equation}\label{eq:L1_loss}
\mathcal L = \frac{1}{mp} ||\hat{Y} - Y||.
\end{equation}

Following the differential chain rule, one could backward-manipulate gradients of parameters in the $i$th layer from the $(i+1)$th layer, which is well recognized as the backpropagation algorithm. Formally, in a vectorized batch style \citep{LeCun:2012},
\begin{align}
\frac{\partial \mathcal L}{\partial Z_{i + 1}}&=f^{\prime}(Z_{i + 1})\frac{\partial \mathcal L}{\partial X_{i+1}},\\
\frac{\partial \mathcal L}{\partial W_{i + 1}}&=X_{i}^T\frac{\partial \mathcal L}{\partial Z_{i+1}},\\
\frac{\partial \mathcal L}{\partial X_{i}}&= W_{i+1}^T \frac{\partial \mathcal L}{\partial Z_{i + 1}}, \\
\frac{\partial \mathcal L}{\partial \bm{b}_{i+1}} &= \sum_j{\bm{\text{row}}_j}\left(\frac{\partial \mathcal L}{\partial Z_{i+1}}\right).
\end{align}
where the operator $\frac{\partial \mathcal L}{\partial \cdot}$ represents elementwise partial derivatives of $\mathcal L$ on corresponding indices, and $f^\prime$ is the derivative of the nonlinear function $f$. The network parameters are then updated by a gradient-based optimizer in each iteration. Here, we adopt Adam \citep{Kingma:2014} as the optimizer, which can accelerate the convergence.

Batch normalization, which was proposed by \cite{Ioffe:2015}, is implemented before every nonlinear layer. Batch normalization is tested to stabilize the distribution among variables, hence it benefits the optimization and accelerates the convergence. It also enables us to use higher learning rates and care less about initialization.

\begin{figure*}
	\centering
	\includegraphics[width=0.45\textwidth]{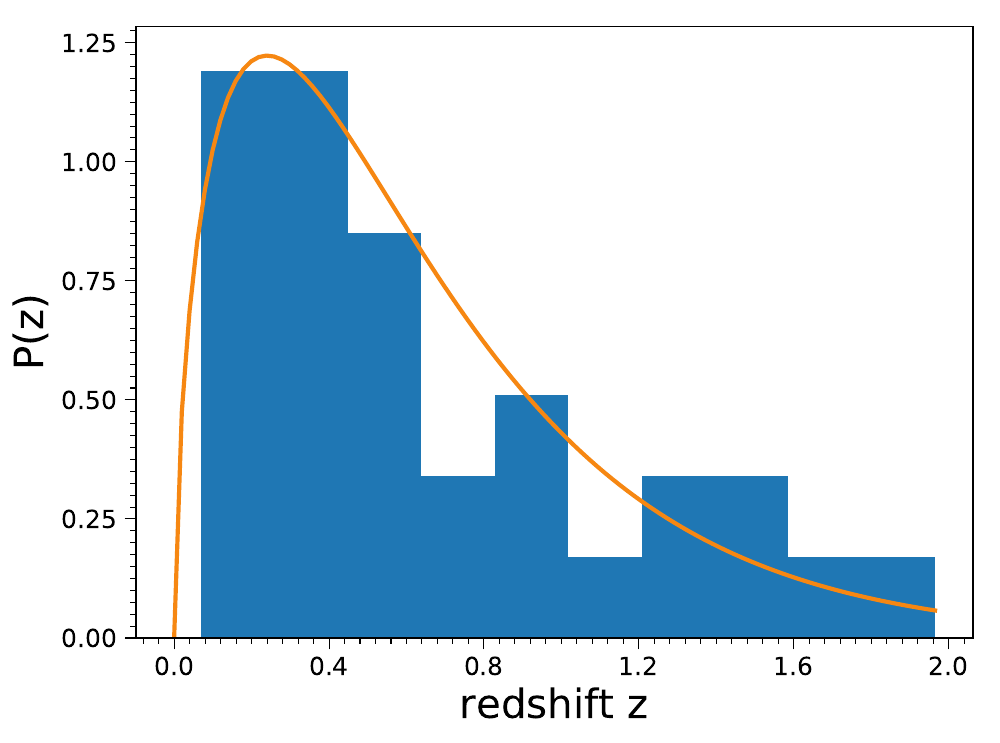}
	\includegraphics[width=0.45\textwidth]{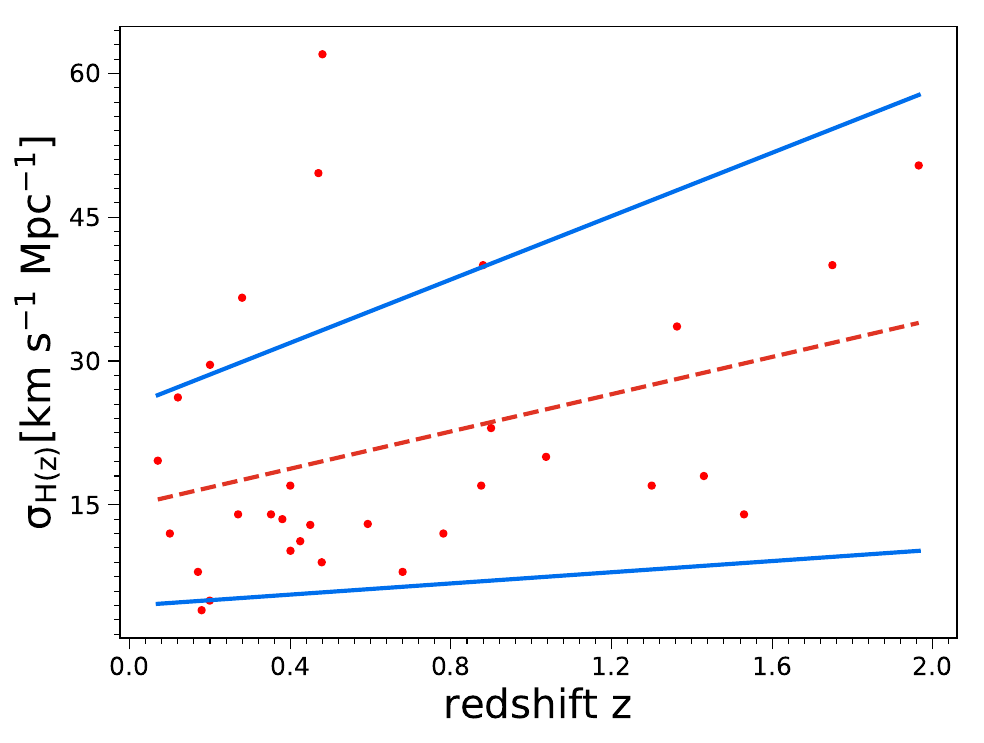}
	\caption{{\it Left:} redshift distribution of the observational $H(z)$. {\it Right:} errors of the observational $H(z)$.}\label{fig:obs_Hz_distribution}
\end{figure*}

\subsection{Simulation of $H(z)$}\label{sec:sim_Hz}

The network model that is used to reconstruct the observational Hubble parameter is optimized by using the mock $H(z)$, which is simulated in the framework of the flat $\Lambda$CDM model using
\begin{equation}\label{equ:fLCDM_Hz}
H(z) = H_0 \sqrt{\Omega_{\rm m}(1+z)^3 + 1-\Omega_{\rm m}}~,
\end{equation}
with the fiducial $H_0=70~\rm km\ s^{-1}\ Mpc^{-1}$ and $\Omega_m=0.3$. We assume the redshift of the observational $H(z)$ (Table \ref{tab:Hz}) subject to a Gamma distribution,
\begin{equation}\label{equ:gammadistribution}
p(x;\alpha,\lambda)=\frac{\lambda^{\alpha}}{\Gamma(\alpha)}x^{\alpha-1}e^{-\lambda x}~,
\end{equation}
where $\alpha$ and $\lambda$ are parameters, and the gamma function is
\begin{equation}
\Gamma (\alpha) = \int_{0}^{\infty} e^{-t}t^{\alpha-1}dt~.
\end{equation}
The distribution of the observational $H(z)$ and the assumed distribution function of the redshift $z$ are shown in the left panel of Figure \ref{fig:obs_Hz_distribution}.

In the right panel of Figure \ref{fig:obs_Hz_distribution}, we plot errors with respect to the redshift $z$. The error of $H(z)$ obviously increases with the redshift. Following \citet{Ma:2011}, we assume that the error of $H(z)$ increases linearly with the redshift. We first fit $\sigma_{H(z)}$ with first degree polynomials and obtain $\sigma_0=9.72z+14.87$ (the red dashed line). Here we assume that $\sigma_0$ is the mean value of $\sigma_{H(z)}$ at a specific redshift. Then, two lines (the blue solid lines) are selected symmetrically around the mean value line to ensure that most data points are in the area between them, and these two lines have functions of $\sigma_-=2.92z+4.46$ and $\sigma_+=16.52z+25.28$. Finally, the error $\tilde{\sigma}(z)$ is generated randomly according to the Gaussian distribution $\mathcal{N}(\sigma_0(z), \varepsilon^2(z))$, where $\varepsilon(z)=(\sigma_+-\sigma_-)/4$ is set to ensure that $\tilde{\sigma}(z)$ falls in the area with a 95\% probability.

The fiducial values of the Hubble parameter $H_{\rm fid}(z)$ that generated using Equation \ref{equ:fLCDM_Hz} are simulated randomly by adding $\Delta H$ that subject to $\mathcal{N}(0, \tilde{\sigma}^2(z))$. Thus, the final simulated Hubble parameter is $H_{\rm sim}(z)=H_{\rm fid}(z)+\Delta H$ with the uncertainty $\tilde{\sigma}(z)$. Therefore, one can simulate samples of Hubble parameter in the flat $\Lambda$CDM model with the assumed distribution of redshift and errors. We note that the mock $H(z)$ is used to optimize the network model, and the assumption of the error of $H(z)$ increasing linearly with the redshift does not affect the reconstruction of the observational $H(z)$; thus, the error model of $H(z)$ is acceptable in our analysis.

\begin{figure*}
	\centering
	\includegraphics[width=0.24\textwidth]{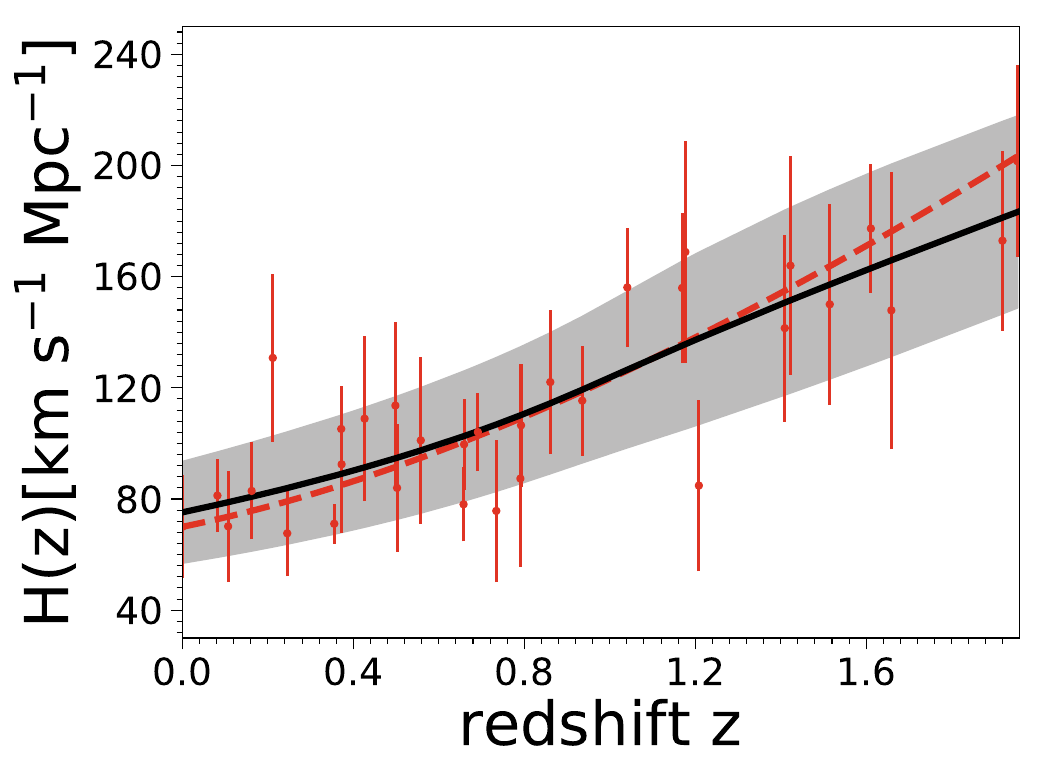}
	\includegraphics[width=0.24\textwidth]{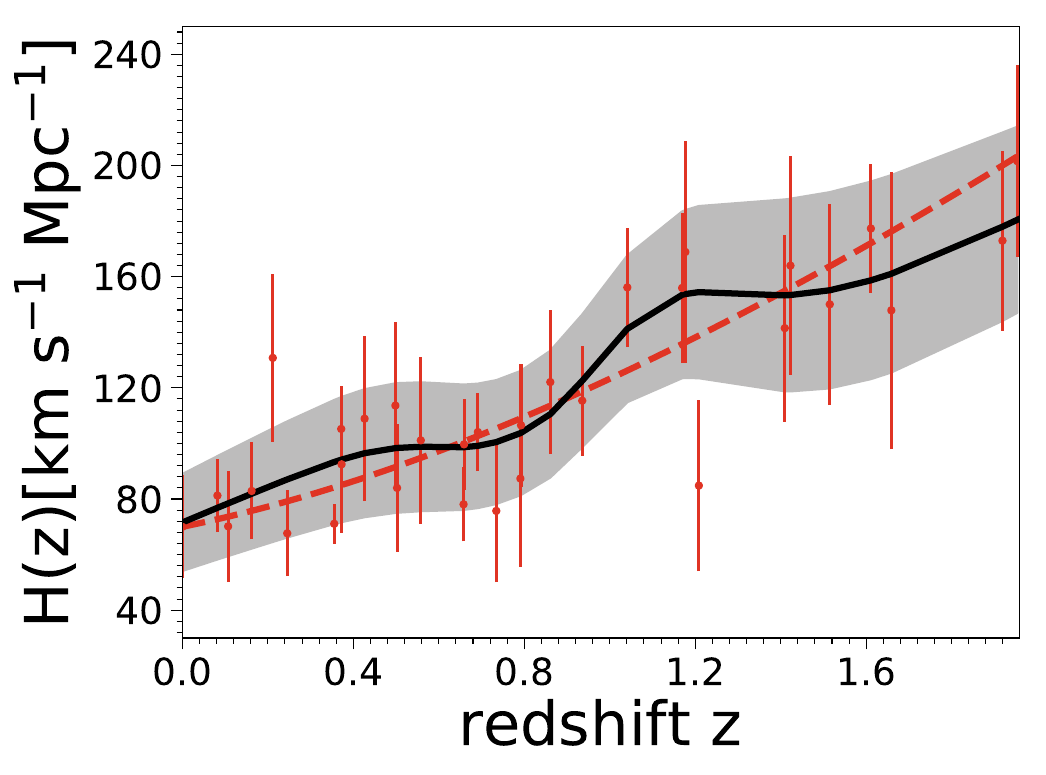}
	\includegraphics[width=0.24\textwidth]{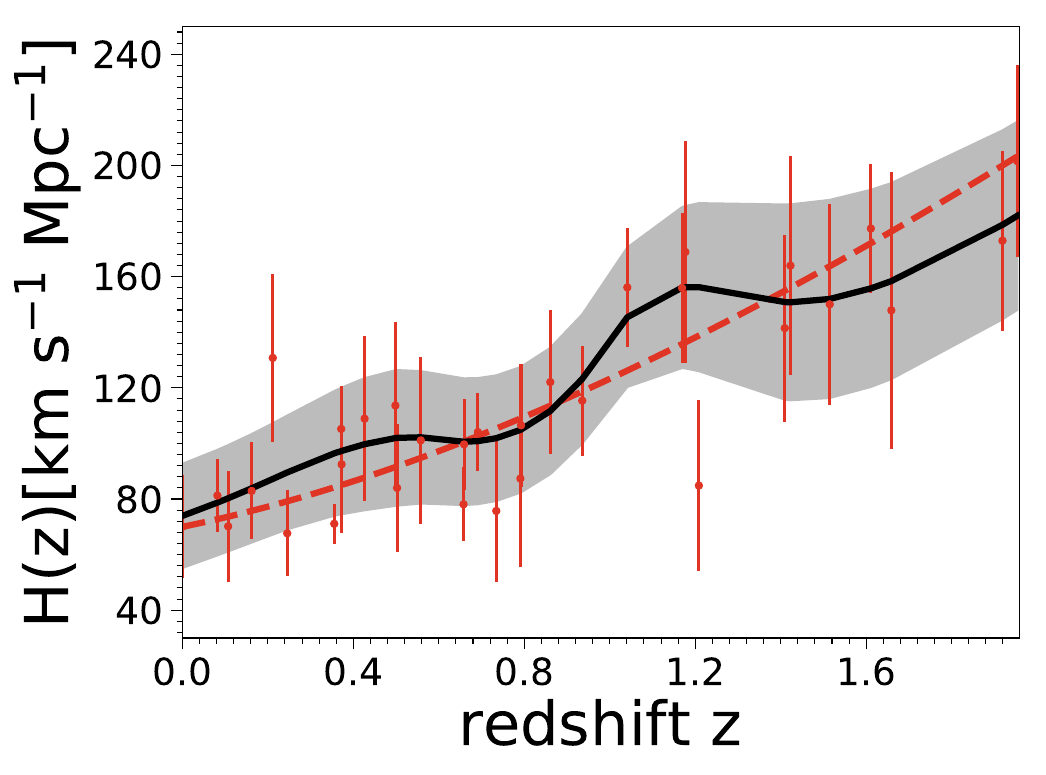}
	\includegraphics[width=0.24\textwidth]{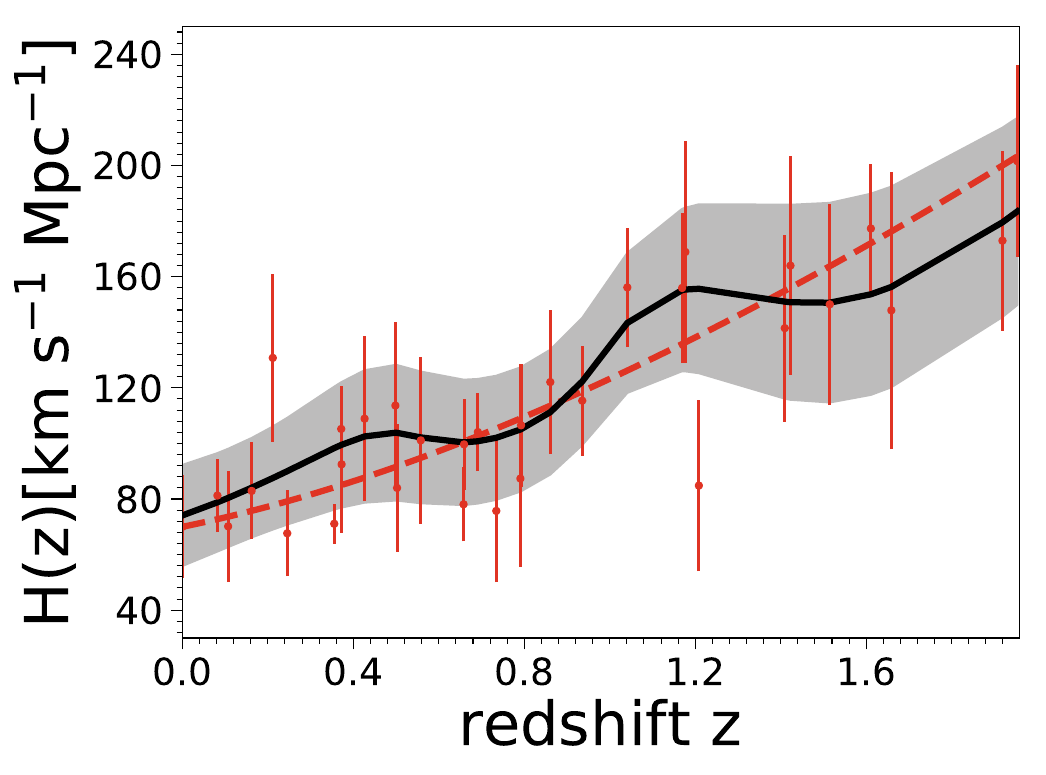}
	\caption{An example of the reconstructed function of $H(z)$ (black lines) and the corresponding $1\sigma$ error (gray areas) with the neural network. The red dots with error bars represent the simulated $H(z)$ data, while the red dashed lines correspond to the fiducial flat $\Lambda$CDM model with $H_0=70~\rm km\ s^{-1}\ Mpc^{-1}$ and $\Omega_{\rm m}=0.3$. From left to right, the network contains 1, 2, 3, and 4 hidden layers, respectively.}\label{fig:mock_Hz}
\end{figure*}

\begin{figure}
	\centering
	\includegraphics[width=0.45\textwidth]{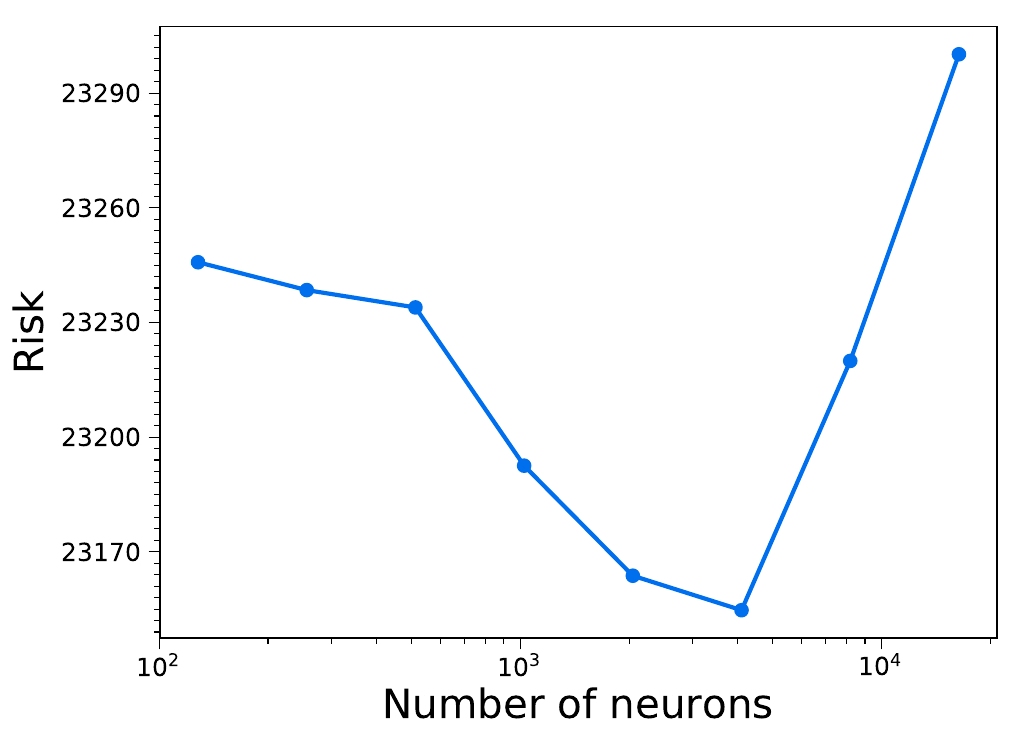}
	\caption{The risk for network models that have one hidden layer. Each network model contains a different number of neurons in its hidden layer.}\label{fig:risk}
\end{figure}

\subsection{Optimize ANN model}\label{sec:optimize_model}

In this section, we illustrate the process of reconstructing functions and find the optimal network model that can be used for the reconstruction of the observational $H(z)$, by using the simulated Hubble parameter. The data used to train the network are simulated according to the redshift distribution of the observational $H(z)$ under the flat $\Lambda$CDM model with the method of section \ref{sec:sim_Hz}. The sample has the same number as that of the observational Hubble parameter, and it is shown in Figure \ref{fig:mock_Hz} (red dots with error bars).

The NN aims to make a mapping from the input data to the output data to construct an approximate function. Specifically, it constructs an approximate function that associates the redshift $z$ with the Hubble parameter $H(z)$ and its uncertainty according to the $H(z)$ data. Thus, the input of the NN is the redshift $z$ and the output is the corresponding $H(z)$ and error $\sigma_{H(z)}$ (see Figure \ref{fig:nn_model}). Parameters of the NN ($W$ and $\bm b$ in Equation \ref{eq:def1}) need to be learned by training the network with data. In supervised learning tasks, the data is commonly divided into three parts: training set, validation set, and test set. The training set is used to train the network model, the validation set is used to tune the hidden parameters (or hyperparameters, such as learning rate, the number of hidden layers, and the number of neurons), and the test set is used to test the accuracy of the NN. However, all of the $H(z)$ data should be used to train the network to construct an approximate function in this task. Thus, there are no validation and test sets in this task, and the evaluation strategy of the NN is different from that of other tasks. Therefore, we present a new strategy to train and evaluate the NN.

The reconstructed function of $H(z)$ should be able to represent an $H(z)$ and its uncertainty at a specific redshift. Thus, an optimal network model should be adopted to learn an approximate function. To illustrate our training and evaluation strategy, we only consider finding the optimal number of hidden layers and the number of neurons in the hidden layer. The initial learning rate is set to 0.01 and decreases with the number of iterations, and the batch size is set to half of the number of the $H(z)$ data. Then, the network model can be trained after multiple iterations by minimizing the loss function of Equation \ref{eq:L1_loss}. In each iteration, a subsample with the number of batch size is randomly selected and is fed to the network; after passing the network, the loss is calculated using the loss function, and it is transmitted backwards to update the weight vectors and bias according to gradient descent. Here we set the number of iterations to be $3\times10^4$, which is large enough to ensure the loss function no longer decreases.

We first estimate the optimal number of hidden layers of the NN using the simulated $H(z)$ data. We train the network with the simulated $H(z)$ sample. The number of hidden layers of the network we consider varies from 1 to 4, and 8 network models are trained with the number of neurons in the range of $[128,16384]$ for each network structure. Thus, 32 network models are trained in total. We note that these 32 network models are trained independently with the same samples of $H(z)$ shown in Figure \ref{fig:mock_Hz}. To choose the optimal number of hidden layers of the network, the statistically correct thing to do is to minimize the {\it risk} \citep{Wasserman:2001}:
\begin{align}
\nonumber\rm risk &= \sum_{i=1}^N{\rm Bias}^2_{i} + \sum_{i=1}^N {\rm Variance}_i\\
&=\sum_{i=1}^N[H(z_i) - \bar{H}(z_i)]^2 + \sum_{i=1}^N\sigma^2(H(z_i))~,
\end{align}
where $N$ is the number of $H(z)$ data points, and $\bar{H}(z)$ is the fiducial value of $H(z)$. We calculate the average of the {\it risk} of 8 models for each network structure and obtain four values of the {\it risk:} 23218, 25326, 26851, 26782 where the number of hidden layers equal 1, 2, 3, and 4, respectively. Thus, the network structure that contains one hidden layer should be chosen as the optimal one. In order to visualize the effect of the number of hidden layers on the $H(z)$ reconstruction, we show an example of reconstructed $H(z)$ with different network structures in Figure \ref{fig:mock_Hz}. The red dashed lines represent the fiducial $\Lambda$CDM model. From left to right, the number of hidden layers of the corresponding network is 1, 2, 3, and 4, respectively. Obviously, with the increase of the hidden layers, the reconstructed $H(z)$ will gradually deviate from the fiducial model.

For further determination of the number of neurons in the hidden layer, we plot the {\it risk} of the 8 network models that contain one hidden layer, shown in Figure \ref{fig:risk}. We can see that the {\it risk} decreases first and then increases with the increase in the number of neurons, and it has a minimal risk when the number of neurons is 4096. Therefore, we choose a network that contains 4096 neurons in the hidden layer as the optimal one, and apply it to the reconstruction of the observational $H(z)$. We note that the hyperparameters of the optimal model are the only things that are adopted in the reconstruction of the observational $H(z)$, and the method of simulating the Hubble parameter of section \ref{sec:sim_Hz} has no effect on the reconstruction of the observational data. To reconstruct functions from other data, the strategy illustrated in this section can also be used to find the optimal network model.

\begin{figure}
	\centering
	\includegraphics[width=0.45\textwidth]{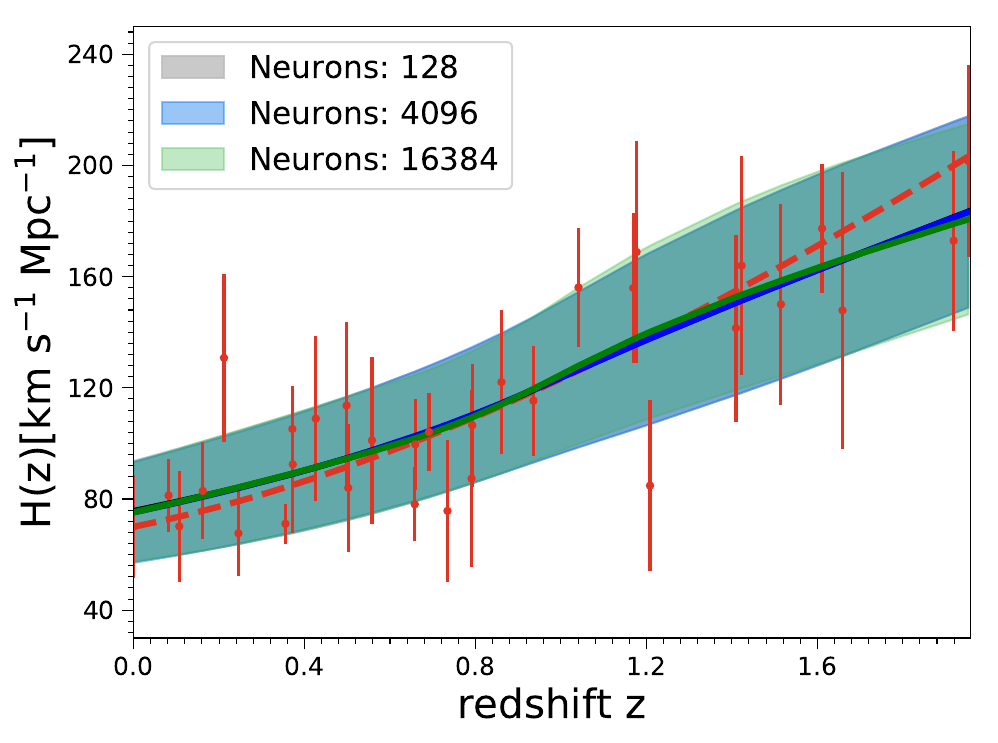}
	\caption{Three reconstructed functions of $H(z)$ and the corresponding $1\sigma$ error with neural networks that have one hidden layer. These functions are reconstructed with networks that have different neurons in the hidden layer. The red dots with error bars represent the simulated $H(z)$ data, while the red dashed lines correspond to the fiducial flat $\Lambda$CDM model with $H_0=70~\rm km\ s^{-1}\ Mpc^{-1}$ and $\Omega_{\rm m}=0.3$.}\label{fig:mock_Hz_layer1}
\end{figure}

We further visualize the effect of the number of neurons in the hidden layer on the reconstruction of $H(z)$, shown in Figure \ref{fig:mock_Hz_layer1}. In this figure, we plot three reconstructed functions of $H(z)$ that are trained with three different network models that have one hidden layer. The number of neurons in the hidden layer of these models is 128, 4096, and 16384, respectively. We can see that these three functions of $H(z)$ are almost the same, which is different from the effect of the number of the hidden layer (see Figure \ref{fig:mock_Hz}). This weak effect of the number of neurons in the hidden layer on the reconstruction of $H(z)$ makes it safe to find the optimal model in 8 kinds of network models with the number of neurons in the hidden layer lying in the range of [128, 16384].

\section{Reconstruction of $H(z)$}\label{sec:reconstruct_Hz}

In this section, we first introduce the Hubble parameter measurements $H(z)$, then utilize the optimal network model selected in section \ref{sec:optimize_model} to reconstruct functions of the observational $H(z)$.

\begin{table}
	\centering \caption{31 CC $H(z)$ measurements obtained from the differential age method. Note. The Hubble paramter obtained from \citet{Ratsimbazafy:2017} is $89\pm23(stat)\pm44(syst) ~\rm km\ s^{-1}\ Mpc^{-1}$, here we consider their total error $89\pm49.6(tot) ~\rm km\ s^{-1}\ Mpc^{-1}$ in our analysis.}\label{tab:Hz}
	\begin{tabular}{ccc}
		\hline
		\hline
		z 		& $H(z)$ (km $\rm s^{-1}$ $\rm Mpc^{-1}$) & References \\
		\hline
		0.09 	&	$69\pm12$		&   \citet{Jimenez:2003} \\
		\hline
		0.17	&	$83\pm8$		&	\\
		0.27	&	$77\pm14$		&	\\
		0.4		&	$95\pm17$		&	\\
		0.9		&	$117\pm23$		&   \citet{Simon:2005} \\
		1.3		&	$168\pm17$		&	\\
		1.43	&	$177\pm18$		&	\\
		1.53	&	$140\pm14$		&	\\
		1.75	&	$202\pm40$		&	\\
		\hline
		0.48	&	$97\pm62$		&   \citet{Stern:2010} \\
		0.88	&	$90\pm40$		&	\\
		\hline
		0.1791	&	$75\pm4$		&	\\
		0.1993	&	$75\pm5$		&	\\
		0.3519	&	$83\pm14$		&	\\
		0.5929	&	$104\pm13$		&	\citet{Moresco:2012} \\
		0.6797	&	$92\pm8$		&	\\
		0.7812	&	$105\pm12$		&	\\
		0.8754	&	$125\pm17$		&	\\
		1.037	&	$154\pm20$		&	\\
		\hline
		0.07	&	$69\pm19.6$		&	\\
		0.12	&	$68.6\pm26.2$	&	\citet{Zhang:2014} \\
		0.2		&	$72.9\pm29.6$	&	\\
		0.28	&	$88.8\pm36.6$	&	\\
		\hline
		1.363	&	$160\pm33.6$	&	\citet{Moresco:2015} \\
		1.965	&	$186.5\pm50.4$	&	\\
		\hline
		0.3802	&	$83\pm13.5$		&	\\
		0.4004	&	$77\pm10.2$		&	\\
		0.4247	&	$87.1\pm11.2$	&	\citet{Moresco:2016} \\
		0.44497	&	$92.8\pm12.9$	&	\\
		0.4783	&	$80.9\pm9$		&	\\
		\hline
		0.47    &   $89\pm49.6$     &  \citet{Ratsimbazafy:2017} \\
		\hline
	\end{tabular}
\end{table}

\subsection{Hubble parameter $H(z)$}

\begin{figure*}
	\centering
	\includegraphics[width=0.32\textwidth]{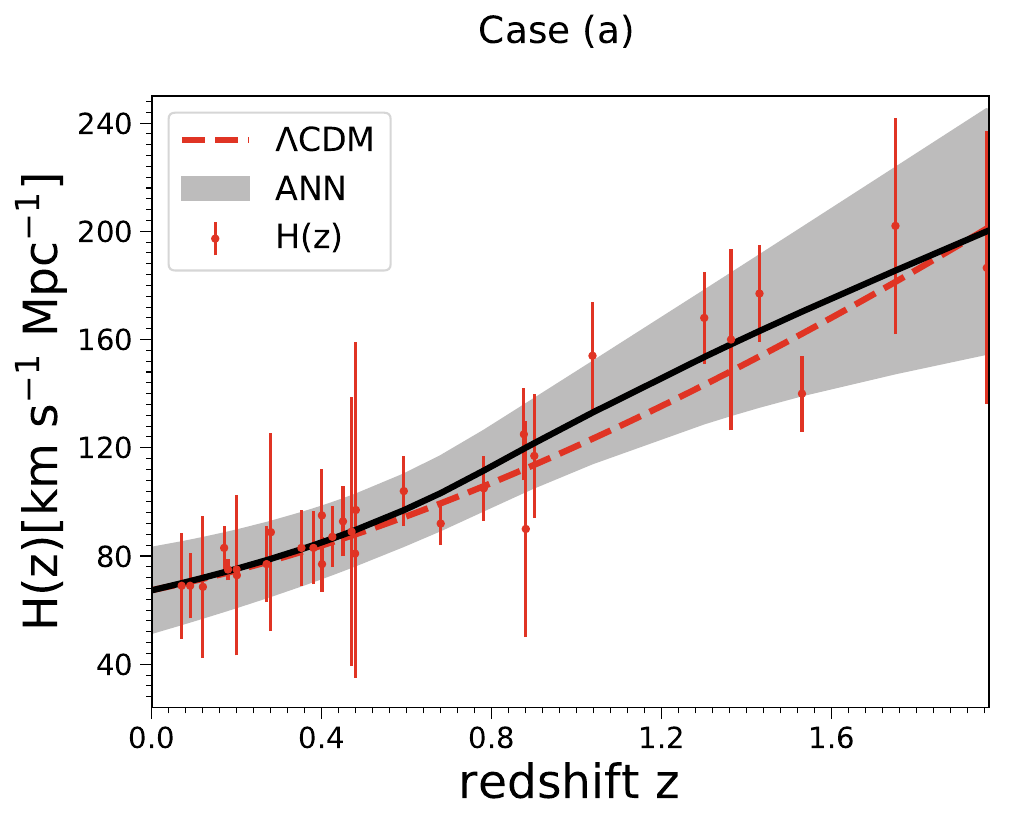}
	\includegraphics[width=0.32\textwidth]{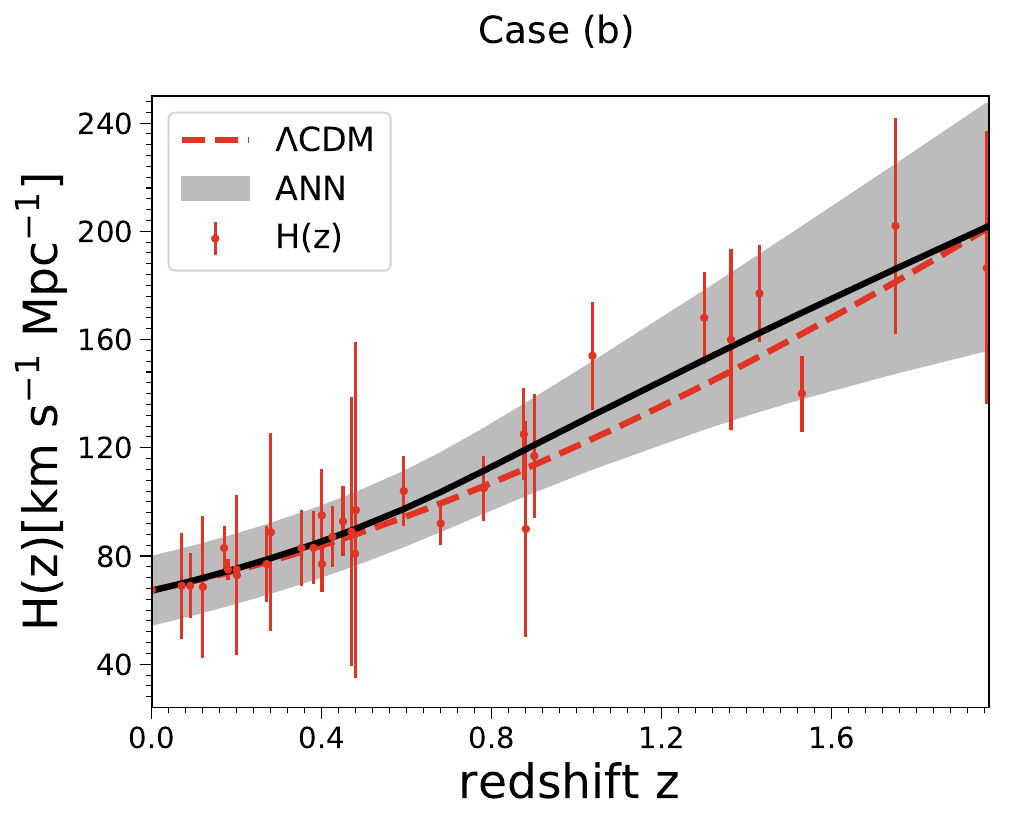}
	\includegraphics[width=0.32\textwidth]{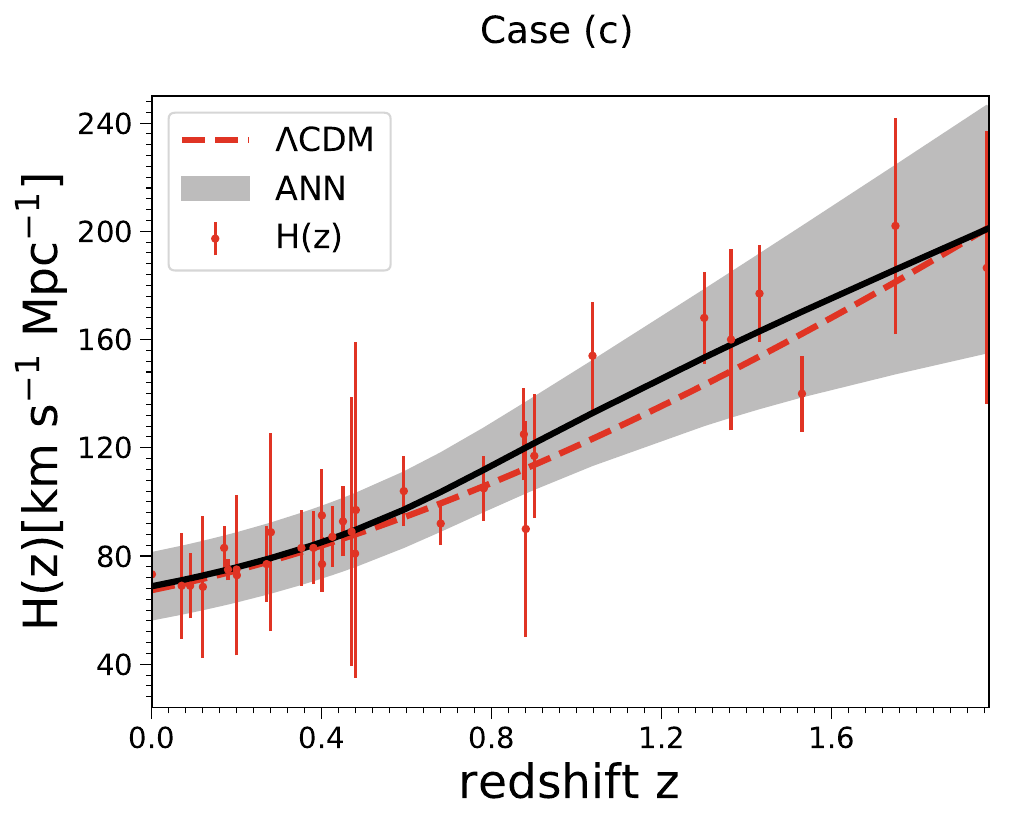}
	\caption{Reconstructed function of $H(z)$ with ANN. The red dots with error bars represent the $H(z)$ data, while the red dashed lines correspond to the best-fit flat $\Lambda$CDM models with $H_0=67.4 ~\rm km\ s^{-1}\ Mpc^{-1}$ and $\Omega_{\rm m}=0.315$ (Planck2018 result). The left panel corresponds to the result of case (a) (no $H_0$ prior), the middle panel represents that of case (b) (with a prior of $H_0=67.4\pm0.5 ~\rm km\ s^{-1}\ Mpc^{-1}$), and the right panel is for case (c) (with a prior of $H_0=73.24\pm1.74 ~\rm km\ s^{-1}\ Mpc^{-1}$).}\label{fig:obs_Hz_rec}
\end{figure*}

The Hubble parameter measurements $H(z)$, which have been used to explore the evolution of the universe and the nature of dark energy, describe the expansion rate of the universe. $H(z)$ can be obtained in two ways. One method to obtain $H(z)$ is based on the detection of the radial BAO features \citep{Gaztanaga:2009,Blake:2012,Samushia:2013}. However, the $H(z)$ data obtained using this method are based on an assumed fiducial cosmological model. Thus, these $H(z)$ data are not considered in our analysis. Another method is to calculate the differential ages of passively evolving galaxies at different redshifts, which provides $H(z)$ measurements that are model-independent \citep{Jimenez:2002}. In this method, a change rate $\Delta z/\Delta t$ can be obtained, then the Hubble parameter $H(z)$ could be written as
\begin{equation}
H(z)\simeq-\frac{1}{1+z}\frac{\Delta z}{\Delta t} ~.
\end{equation}
This method is usually called the cosmic chronometers (CCs), and the $H(z)$ data based on this method are referred to as CC $H(z)$. On the basis of the CC $H(z)$ data used in \citet{Wang:2017}, we add another new $H(z)$ measurement taken from \citet{Ratsimbazafy:2017} to achieve our model-independent analysis. Hence, the $H(z)$ sample has 31 data points totally within the redshift range of [0.07, 1.965], which are correctly summarized in Table \ref{tab:Hz}. Note that the $H(z)$ taken from \citet{Ratsimbazafy:2017} is $89\pm23(stat)\pm44(syst) ~\rm km\ s^{-1}\ Mpc^{-1}$; thus, the $H(z)$ with total error $89\pm49.6(tot) ~\rm km\ s^{-1}\ Mpc^{-1}$ is considered in our analysis.

\subsection{Functions of CC $H(z)$}\label{sec:reconstruct_CCHz}

The minimum redshift of the CC $H(z)$ is 0.07, which is larger than most of the current SN Ia data. Thus, if we want to explore a lower redshift universe with the Hubble parameter, one possible way is to extend the reconstructed $H(z)$ function to a lower redshift. However, it should be noted that this extension is completely an approximation, which may introduce bias when having few $H(z)$ data in the vicinity of the redshift interval. Therefore, we consider a prior of the Hubble constant $H_0$ in the reconstruction of $H(z)$ to make the reconstructed function of $H(z)$ more reliable. We adopt two recent measurements of $H_0$ in the reconstruction of $H(z)$: $H_0=67.4\pm0.5 ~\rm km\ s^{-1}\ Mpc^{-1}$ with 0.7\% uncertainty \citep
{Planck2018:VI}, and $H_0=73.24\pm1.74 ~\rm km\ s^{-1}\ Mpc^{-1}$ with 2.4\% uncertainty \citep{Riess:2016}. Further, for comparison, we also reconstruct $H(z)$ with no $H_0$ prior. Thus, there are 3 cases when reconstructing $H(z)$:
\begin{itemize}
	\item [(a)] with no $H_0$ prior;
	\item [(b)] with a prior of $H_0=67.4\pm0.5 ~\rm km\ s^{-1}\ Mpc^{-1}$; and
	\item [(c)] with a prior of $H_0=73.24\pm1.74 ~\rm km\ s^{-1}\ Mpc^{-1}$.
\end{itemize}
For case (a), the training set has 31 observational Hubble parameters listed in Table \ref{tab:Hz}, and for cases (b) and (c), the training set contains an additional data point at the redshift $z=0$, thus, the training set has 32 data points for cases (b) and (c).

\begin{figure*}
	\centering
	\includegraphics[width=0.32\textwidth]{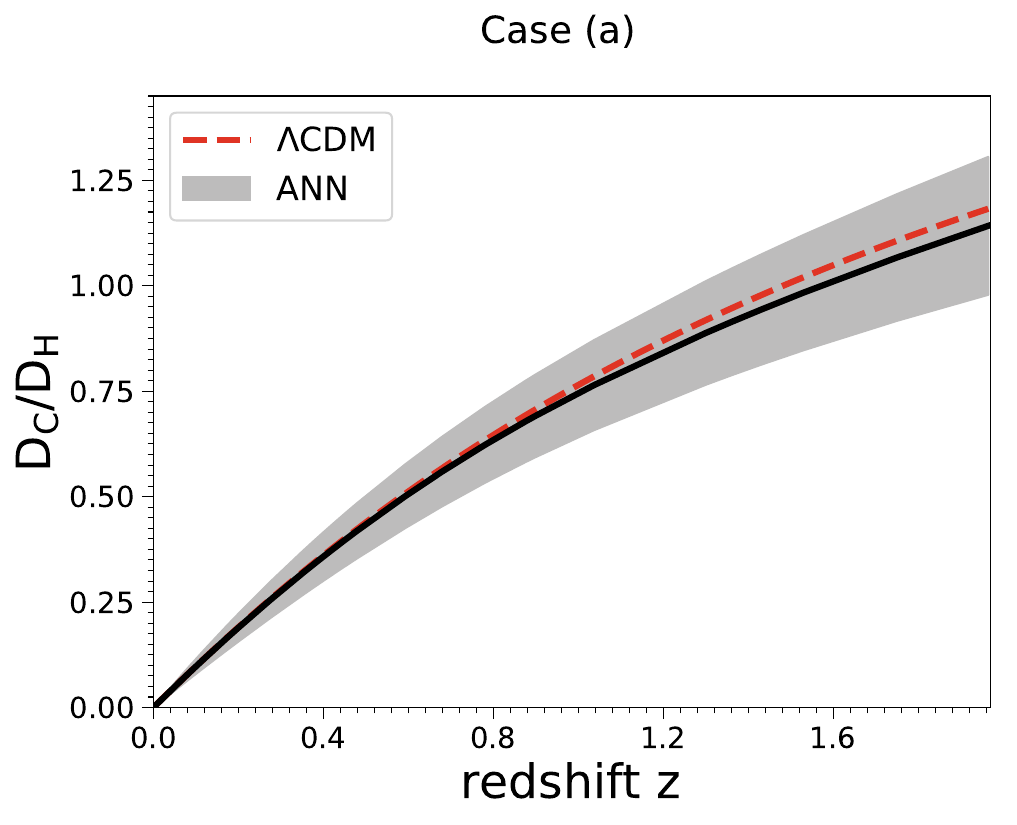}
	\includegraphics[width=0.32\textwidth]{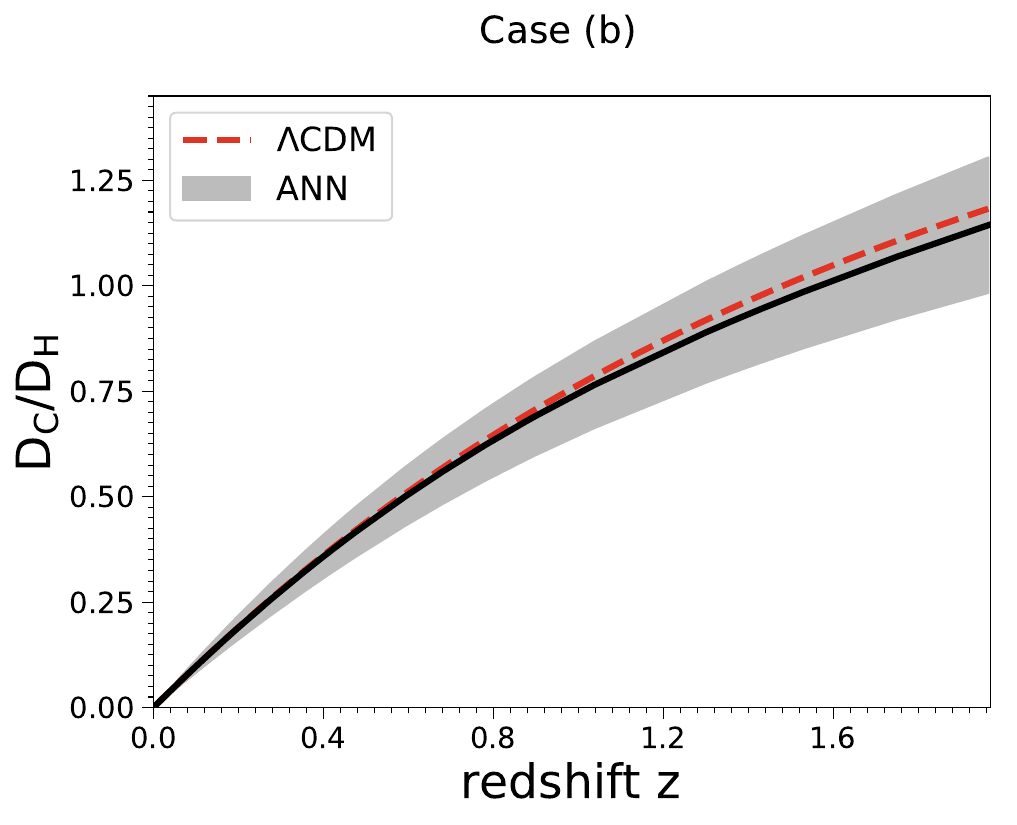}
	\includegraphics[width=0.32\textwidth]{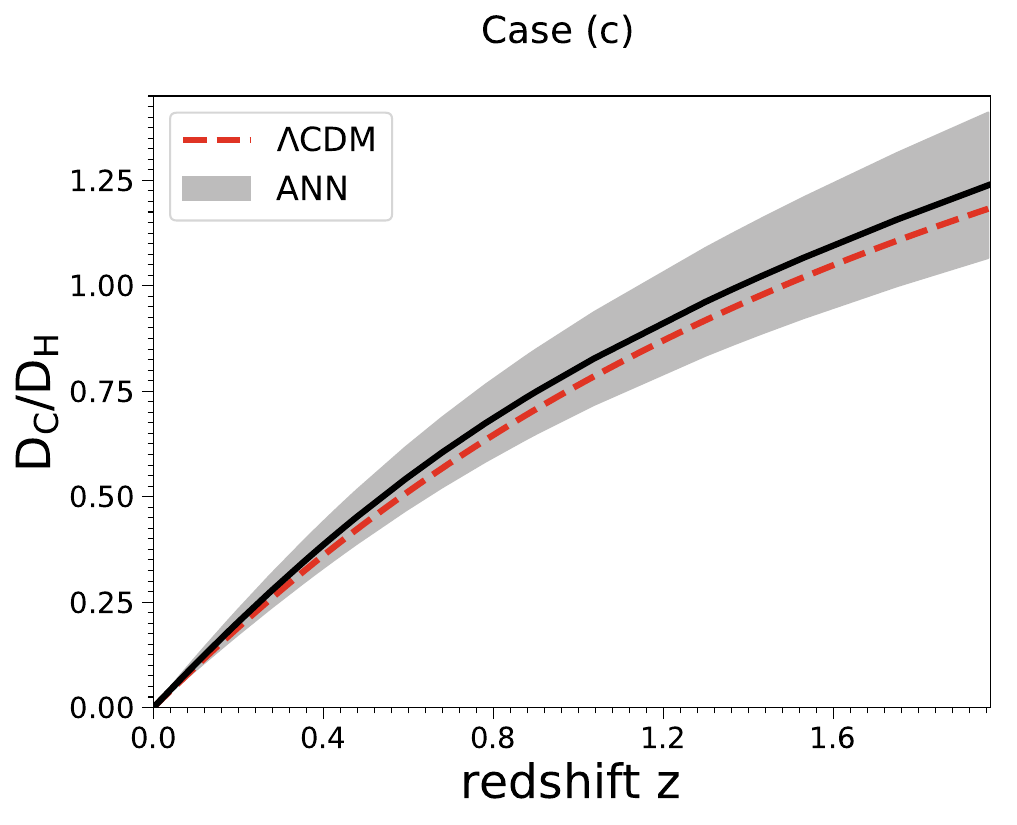}
	\caption{Reconstructed function of $\rm D_C/D_H$ with ANN, where $\rm D_C$ is obtained by integrating functions of $H(z)$ in Figure \ref{fig:obs_Hz_rec} using Equation \ref{equ:comoving}, and $D_{\rm H}=cH_0^{-1}$. The red dashed lines correspond to the best-fit flat $\Lambda$CDM models with $H_0=67.4 ~\rm km\ s^{-1}\ Mpc^{-1}$ and $\Omega_{\rm m}=0.315$ (Planck2018 result).}\label{fig:obs_Hz_rec_DC}
\end{figure*}

Using the optimal network model obtained in Section \ref{sec:optimize_model}, we reconstruct functions of $H(z)$ by training the network for the three cases of the $H(z)$ sample. After training the NN, one can feed a sequence of redshifts to the network model and obtain a series of Hubble parameter with errors. Thus, the output Hubble parameters and corresponding errors, as well as the input redshift sequence, constitute a function of $H(z)$. The reconstructed functions of $H(z)$ for the three cases are shown in Figure \ref{fig:obs_Hz_rec}. The red dots with error bars represent the observational $H(z)$, and the red dashed lines are the flat $\Lambda$CDM model with $H_0=67.4 ~\rm km\ s^{-1}\ Mpc^{-1}$ and $\Omega_{\rm m}=0.315$ (Planck2018 result; \citet{Planck2018:VI}). The black lines and gray areas are the best values and $1\sigma$ errors of the reconstructed functions of $H(z)$. Obviously, the reconstructed functions
are consistent with those of the flat $\Lambda$CDM model within a $1\sigma$ confidence level for all these three cases. Moreover, we can see that the functions of $H(z)$ for these three cases are very similar to each other. Obviously, they are consistent with each other within a 1$\sigma$ confidence level. For the best values of the reconstructed $H(z)$ (the black solid lines in Figure \ref{fig:obs_Hz_rec}), the relative deviation of case (b) with respect to case (a) is $<0.8\%$, and the relative deviation of case (c) with respect to case (a) is $<1.6\%$. We note that, for case (a), the reconstructed Hubble constant is
\begin{equation}\label{equ:H0_ANN}
H_0=67.33\pm15.74~ \rm km~ s^{-1} ~Mpc^{-1}~,
\end{equation}
where the best-fit value is similar to the latest {\it Planck} CMB result: $H_0=67.4\pm 0.5~\rm km~ s^{-1} ~Mpc^{-1}$. Then, we obtain the total line-of-sight comoving distance $D_{C}$ \citep{Hogg:1999} by using
\begin{equation}\label{equ:comoving}
D_{C}=c\int_{0}^{z}\frac{dz'}{H(z')} ~.
\end{equation}
The error of $D_C$ is obtained by integrating the error of the $H(z)$ functions. The corresponding reconstructed $D_C/D_{\rm H}$ are shown in Figure \ref{fig:obs_Hz_rec_DC}, where $D_{\rm H}=cH_0^{-1}$.

To quantify the reliability of the reconstructed functions of $H(z)$, we fit the flat $\Lambda$CDM model using the data generated by these functions of $H(z)$ by comparing the distance modulus. The NN learns complex relationships between the redshift and the corresponding Hubble parameter and its error. Specifically, the black lines in Figure \ref{fig:obs_Hz_rec} represent the evolution of the Hubble parameter with the redshift, and the gray areas refer to the distribution of the errors of the Hubble parameter with the redshift. Thus, the reconstructed error of $H(z)$ is entirely determined by the observational data standing in for the error level of the Hubble parameter at a specific redshift. It should be noted that any number of Hubble parameters can be obtained by feeding a sequence of redshifts to the network model. Therefore, this will lead to the cosmological parameters being constrained to an arbitrary precision when using much more samples than the observational data, which is unreasonable. To avoid this, the covariance between any two different points should be considered. However, the covariance cannot be generated by the ANN model of Figure \ref{fig:nn_model}, and we will discuss this issue in Section \ref{sec:discussion_cov}. Therefore, in order to mitigate the effect of covariance, we generate the same number of $H(z)$ as the training set from the reconstructed function of $H(z)$ and assume the data points are independent. Thus, the approximate $\chi^2$ here takes the form of
\begin{equation}\label{equ:chi2}
\chi^2(H_0, \Omega_{\rm m}) = \sum_{i}\frac{\left[\mu_{\rm th}(z_i;H_0, \Omega_{\rm m})-\mu_{H}(z_i)\right]^{2}}{\sigma^2_{\mu_H,i}}~,
\end{equation}
where 
\begin{align}\label{equ:mu_LCDM}
\mu_H&=5\log\frac{D_L}{\rm Mpc}+25 ~, & D_L&=(1+z)D_C ~,
\end{align}
and the corresponding errors can be propagated by using
\begin{align}
\sigma_{\mu_H}&=\frac{5}{\ln10}\frac{\sigma_{D_L}}{D_L}~,  & \sigma_{D_L}&= (1+z)\sigma_{D_C} ~.
\end{align}
The Hubble constant $H_0$ is needed for the integral of Equation \ref{equ:comoving}; thus, for case (a), we adopt the Hubble constant reconstructed using the ANN method (Equation \ref{equ:H0_ANN}).

\begin{figure*}
	\centering
	\includegraphics[width=0.32\textwidth]{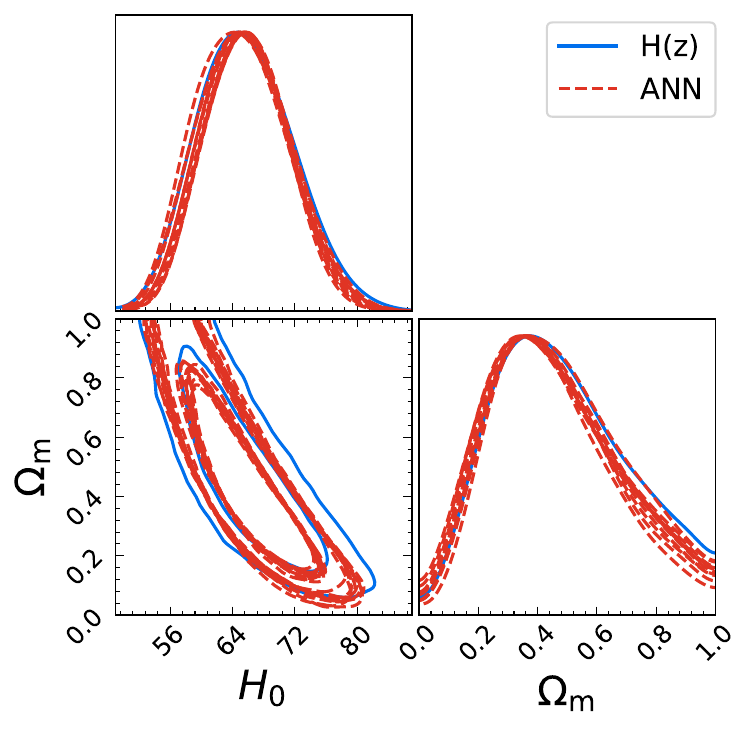}
	\includegraphics[width=0.32\textwidth]{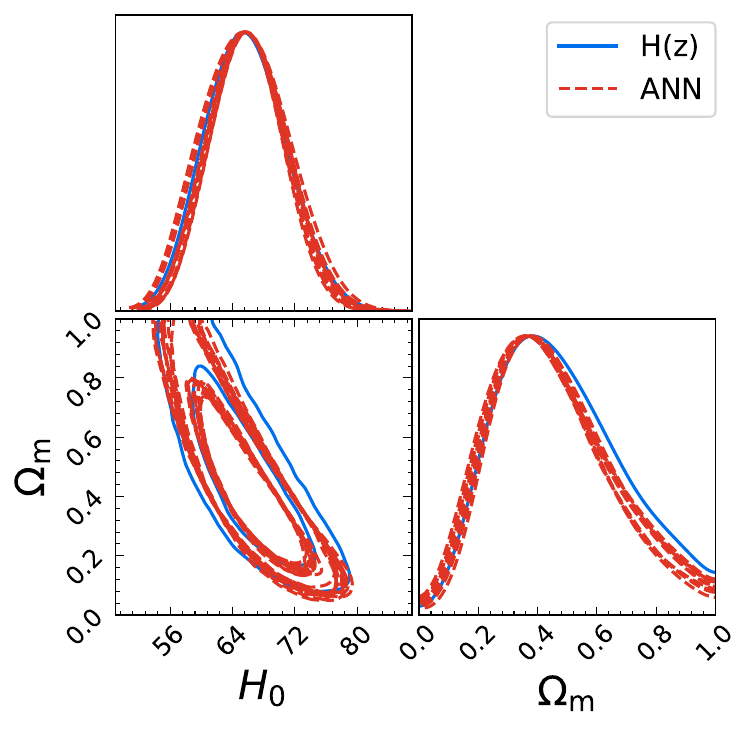}
	\includegraphics[width=0.32\textwidth]{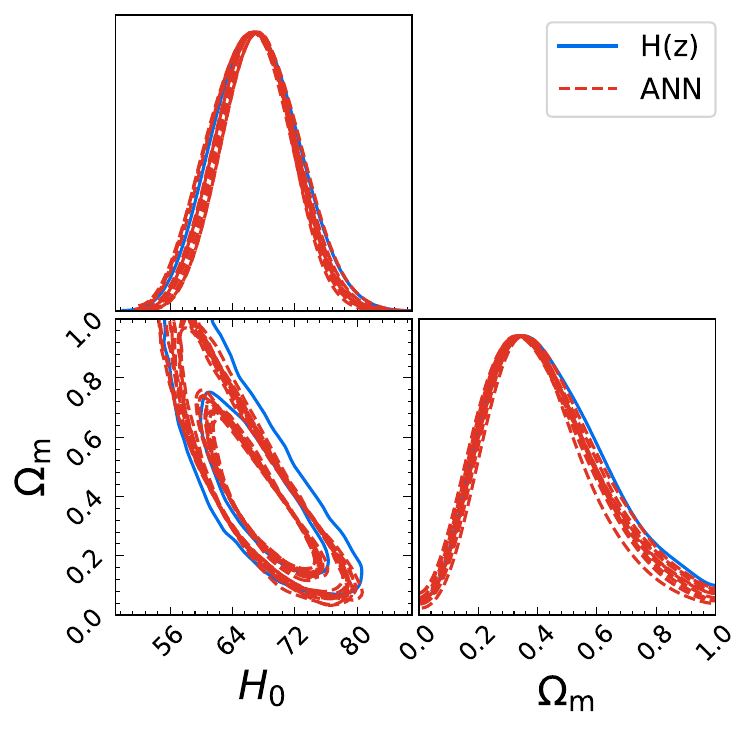}
	\caption{One-dimensional and two-dimensional marginalized distributions with $1\sigma$ and $2\sigma$ contours of $H_0$ and $\Omega_{\rm m}$ constrained from Hubble parameter $H(z)$. The blue solid lines show the results of fitting the $\Lambda$CDM model with $H(z)$ data directly, and the red dashed lines refer to the results of fitting the $\Lambda$CDM model with the reconstructed $H(z)$ using the ANN method. The left panel shows the result of case (a) (no $H_0$ prior), the middle panel refers to the result of case (b) (with a prior of $H_0=67.4\pm0.5 ~\rm km\ s^{-1}\ Mpc^{-1}$), while the right panel stands for case (c) (with a prior of $H_0=73.24\pm1.74 ~\rm km\ s^{-1}\ Mpc^{-1}$). See the text for details.}\label{fig:obs_Hz_LCDM}
\end{figure*}

We first integrate the observational $H(z)$ to obtain the corresponding distance modulus and then fit the $\Lambda$CDM model using the Markov Chain Monte Carlo method by minimizing the $\chi^2$ of Equation \ref{equ:chi2}. The results are shown in Figure \ref{fig:obs_Hz_LCDM} (blue lines), and these results are taken as the ground truth. Then, we simulate 10 sets of samples of the redshift $z$ randomly according to the redshift distribution of the observational $H(z)$ (Equation \ref{equ:gammadistribution}), where the sample has the same number of redshifts as the training set. The corresponding $H(z)$ values and errors can be obtained from the reconstructed functions of $H(z)$. Thus, 10 simulated samples of $H(z)$ can be obtained for cases (a), (b), and (c), respectively. We fit the $\Lambda$CDM model using these samples and obtain the distributions of the parameters, shown in Figure \ref{fig:obs_Hz_LCDM} (red dashed lines). These results are almost the same as the ground truth (blue solid lines) that obtained using the observational $H(z)$ data. Thus, this may indicate the reliability of the functions of $H(z)$ reconstructed using the NN. Moreover, we note that the results of the ANN method and the ground truth are similar for all three cases. Therefore, the ANN method is not sensitive to the prior of the Hubble constant.

\section{Reconstruction of $D_L(z)$}\label{sec:reconstruct_DL}

In the cosmology literature, except for the Hubble parameter, the distance-redshift relations are also frequently reconstructed from SNe Ia \citep{Seikel:2012a,Yahya:2014,Yang:2015,Wang:2019} and GW measurements \citep{Liao:2019}. Here we test the feasibility of the ANN method in reconstructing the luminosity distance obtained from SNe Ia. The data used here are from Union2.1 \citep{Suzuki:2012}, which contains 580 SNe Ia in the redshift range of [0.015, 1.414]. The distance modulus of Union2.1 SNe Ia is
\begin{equation}
\begin{split}
\mu_{SNe}(\alpha, \beta, \delta, M_\mathrm{B})=m_\mathrm{B}^*-M_\mathrm{B}+\alpha\times x_1-\beta\times c\\+\delta \cdot P(m^{\mathrm{true}}_*<m_*^{\mathrm{threshold}}),
\end{split}
\end{equation}
where $M_B$ is the absolute $B$-band magnitude of SNe Ia; and $\alpha$, $\beta$, and $\delta$ are nuisance parameters of SNe Ia. We only want to test the feasibility of ANN in reconstructing functions with SN Ia data; thus, $\alpha$, $\beta$, $\delta$, and $M_B$ are set to 0.122, 2.466, -0.036, and -19.318, respectively \citep{Suzuki:2012}. The error of the distance modulus is
\begin{equation}
\sigma_{\mu_{SNe}} = \sqrt{\sigma^2_{m_\mathrm{B}^*} + (\alpha\sigma_{x_1})^2 + (\beta\sigma_{c})^2}.
\end{equation}
Then, the luminosity distance can be obtained by using
\begin{align}
D_L^{SNe} = 10^{(\mu_{SNe}-25)/5},
\end{align}
and the corresponding error is
\begin{equation}
\sigma_{D_L^{SNe}} = \frac{10^{(\mu_{SNe}-25)/5}\cdot\ln10}{5}\cdot\sigma_{\mu_{SNe}}.
\end{equation}

\begin{figure}
	\centering
	\includegraphics[width=0.45\textwidth]{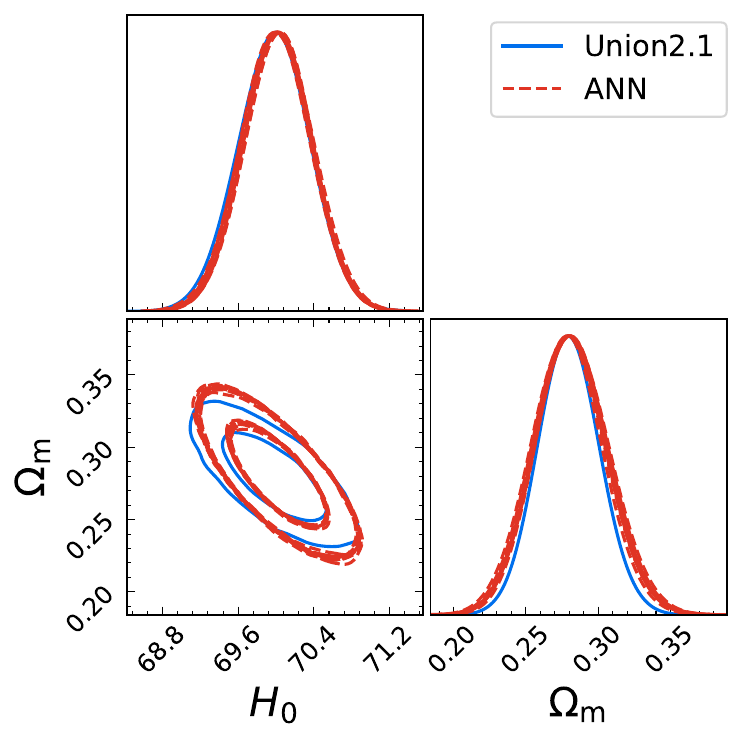}
	\caption{One-dimensional and two-dimensional marginalized distributions with $1\sigma$ and $2\sigma$ contours of $H_0$ and $\Omega_m$ constrained from Union2.1 SNe Ia. The blue solid lines show the results of fitting the $\Lambda$CDM model with Union2.1 SNe Ia directly, and the red dashed lines refer to the results of fitting the $\Lambda$CDM model with the reconstructed SNe Ia using the ANN method.}\label{fig:union2.1_LCDM}
\end{figure}

\begin{figure}
	\centering
	\includegraphics[width=0.45\textwidth]{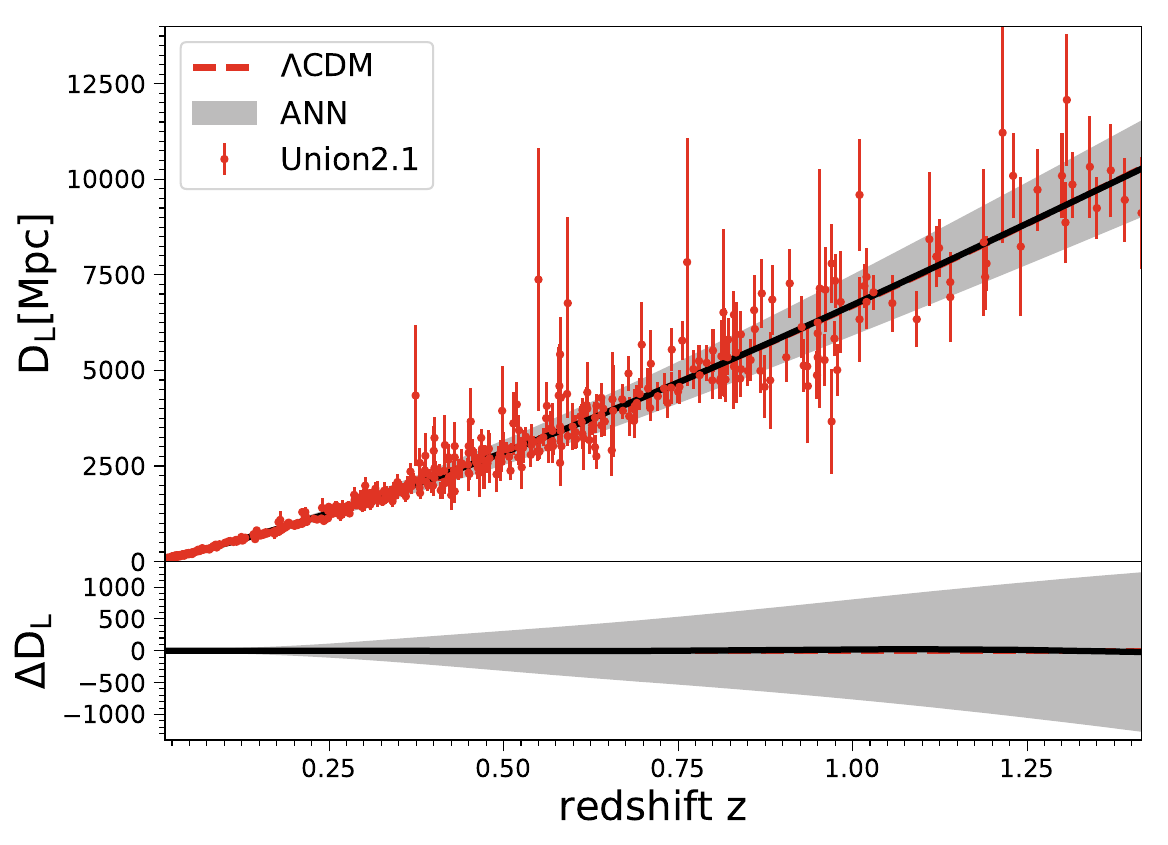}
	\caption{Reconstructed functions of $D_L(z)$ using ANN. In the upper panel, the red dots with error bars represent the $D_L(z)$ of Union2.1 SNe Ia, and the red dashed line refer to the best-fit flat $\Lambda$CDM model of Union2.1 SNe Ia with $H_0=68.984$ and $\Omega_m=0.280$ (Equation \ref{equ:union2.1_bestfit}). Residuals with respect to this model are shown in the lower panel.}\label{fig:union2.1_DL}
\end{figure}

The luminosity distance of Union2.1 is shown in Figure \ref{fig:union2.1_DL} (the red dots with error bars). We first fit the flat $\Lambda$CDM model using the Union2.1 SNe by minimizing the $\chi^2$
\begin{equation}\label{equ:chi2_union2.1}
\chi^2(H_0, \Omega_{\rm m}) = \sum_{i}\frac{\left[D_L^{\rm th}(z_i;H_0, \Omega_{\rm m})-D_L^{SNe}(z_i)\right]^{2}}{\sigma^2_{D_{L}^{SNe},i}}~.
\end{equation}
Note that the absolute magnitude $M_B$ of SNe Ia is fixed; thus, the Hubble constant $H_0$ can be constrained by the SN Ia data. The constraints on $H_0$ and $\Omega_m$ are
\begin{align}\label{equ:union2.1_bestfit}
H_0&=69.984\pm0.347, & \Omega_m&=0.280\pm0.020~,
\end{align}
and one-dimensional and two-dimensional distributions of the parameters are shown in Figure \ref{fig:union2.1_LCDM} (the blue solid lines).

Then, using the method illustrated in section \ref{sec:optimize_model}, we obtain the optimal network model for the reconstruction of $D_L(z)$. The optimal model for reconstructing $D_L(z)$ has one hidden layer, with 4096 neurons in the hidden layer. We note that batch normalization is not used in this model. The reconstructed function of $D_L$ is shown in the upper panel of Figure \ref{fig:union2.1_DL}, where the solid black line with the gray area represents the function of $D_L$ and the corresponding $1\sigma$ error, and the red dashed line stands for the best-fit flat $\Lambda$CDM model of Union2.1 SNe Ia (Equation \ref{equ:union2.1_bestfit}). Residuals with respect to this model are shown in the lower panel. We can see that the function of $D_L$ reconstructed with the ANN method completely coincides with the best-fit flat $\Lambda$CDM model. This indicates that the ANN is capable of reconstructing functions for the distance-redshift relation.

In order to test the reliability of the reconstructed function of $D_L(z)$, we fit the flat $\Lambda$CDM model using 10 sets of samples of SNe Ia generated randomly from the function of $D_L(z)$ according to the redshift distribution of Union2.1 SNe Ia. These samples has the same number of SNe Ia as the Union2.1 SNe Ia. Note that the covariance between any two different points is not considered; thus, the data points are assumed to be independent, and Equation \ref{equ:chi2_union2.1} is an approximate form for these samples. The mean values of parameters for these 10 sets of samples are
\begin{align}\label{equ:union2.1_bestfit_ANN}
H_0&=70.015\pm0.334, & \Omega_m&=0.280\pm0.023~,
\end{align}
and the distributions of parameters are shown in Figure \ref{fig:union2.1_LCDM} (the red dashed lines). This constraint on parameters is almost the same as that obtained from the Union2.1 SNe Ia directly, which further indicates the reliability of the reconstructed function of $D_L(z)$ and makes the ANN a promising method for future cosmological research.

\section{Comparing with Other Networks}\label{sec:compare_other_NN}

For comparison, we also reconstruct functions of $H(z)$ with other NNs. Specifically, we consider the Elman Recurrent Neural Network (RNN, \citet{Elman:1990}), Long Short Term Memory (LSTM, \citet{Hochreiter:1997}), and Gated Recurrent Unit (GRU, \citet{Cho:2014}). In our analysis, the network models built in PyTorch are adopted, and all the processes are carried out on the simulated Hubble parameter.

\begin{figure}
	\centering
	\includegraphics[width=0.45\textwidth]{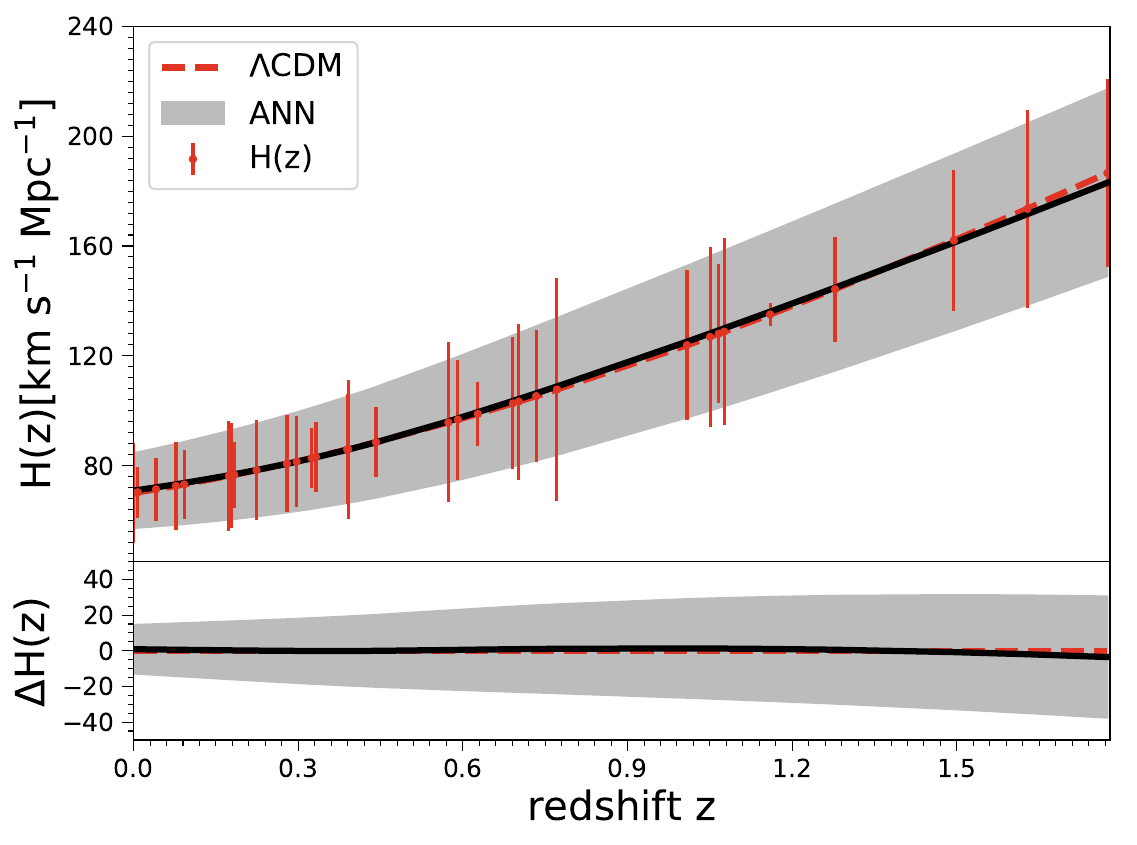}
	\caption{Reconstructed functions of $H(z)$ with $1\sigma$ errors using ANN. In the upper panel, the red dots with error bars represent the simulated $H(z)$, and the red dashed line refers to the fiducial flat $\Lambda$CDM model with $H_0=70~\rm km\ s^{-1}\ Mpc^{-1}$ and $\Omega_m=0.3$. Residuals with respect to this model are shown in the lower panel.}\label{fig:sim_Hz_NoRandom_rec}
\end{figure}

\subsection{ANN}

Using the method of section \ref{sec:sim_Hz}, we firstly simulate a set of samples of the Hubble parameter that has the same number as the observational data, shown in the upper panel of Figure \ref{fig:sim_Hz_NoRandom_rec} (the red dots with error bars). Note that the best values of $H(z)$ are on the fiducial cosmological model. Then, we reconstruct functions of $H(z)$ with the ANN method. The reconstructed functions with $1\sigma$ errors are shown in the upper panel of Figure \ref{fig:sim_Hz_NoRandom_rec}, where the black solid line with the gray area represents the reconstructed function of $H(z)$. In the lower panel of this figure, we show the residual with respect to the fiducial model. The red dashed lines in this figure stand for the fiducial $\Lambda$CDM model. Obviously, the function of $H(z)$ reconstructed with the ANN method coincides completely with the fiducial one.

\begin{figure}
	\centering
	\includegraphics[width=0.45\textwidth]{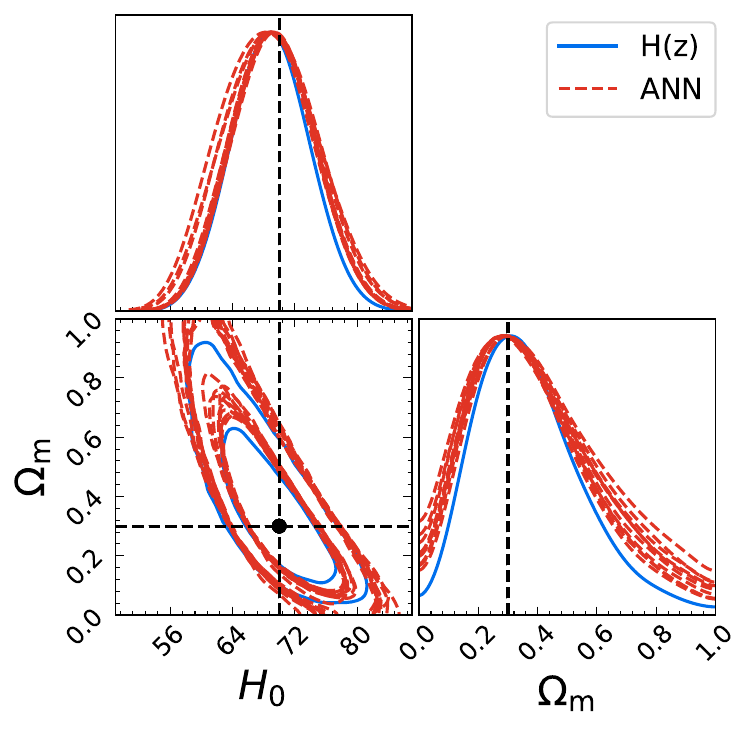}
	\caption{The same as Figure \ref{fig:obs_Hz_LCDM}, except now using the simulated $H(z)$. The black dot stands for the fiducial values of the parameters.}\label{fig:sim_Hz_NoRandom_LCDM}
\end{figure}

We further constrain the parameters of the $\Lambda$CDM model using the same procedure in section \ref{sec:reconstruct_CCHz} with 10 sets of samples generated randomly by the reconstructed function of $H(z)$. One-dimensional and two-dimensional distributions of $H_0$ and $\Omega_m$ constrained from $H(z)$ are shown in Figure \ref{fig:sim_Hz_NoRandom_LCDM}, where the blue lines represent the result constrained from the mock $H(z)$ directly, and the values of the parameter are
\begin{align}
H_0&=68.885\pm4.626, & \Omega_m&=0.351\pm0.174~.
\end{align}
Here, we take this result as the ground truth. The red dashed lines correspond to the results constrained from the $H(z)$ data generated randomly according to the redshift distribution of the observational $H(z)$ from the functions of $H(z)$ reconstructed by ANN. The mean values of $H_0$ and $\Omega_m$ for these 10 sets of samples are
\begin{align}
\nonumber H_0&=69.067\pm5.289, & \Omega_m&=0.361\pm0.221 ~.
\end{align}
Obviously, the result is consistent with the ground truth, and the values of the parameters of the fiducial model are covered by this result within the $1\sigma$ confidence level.

\subsection{RNN, LSTM, and GRU}

\begin{figure}
	\centering
	\includegraphics[width=0.45\textwidth]{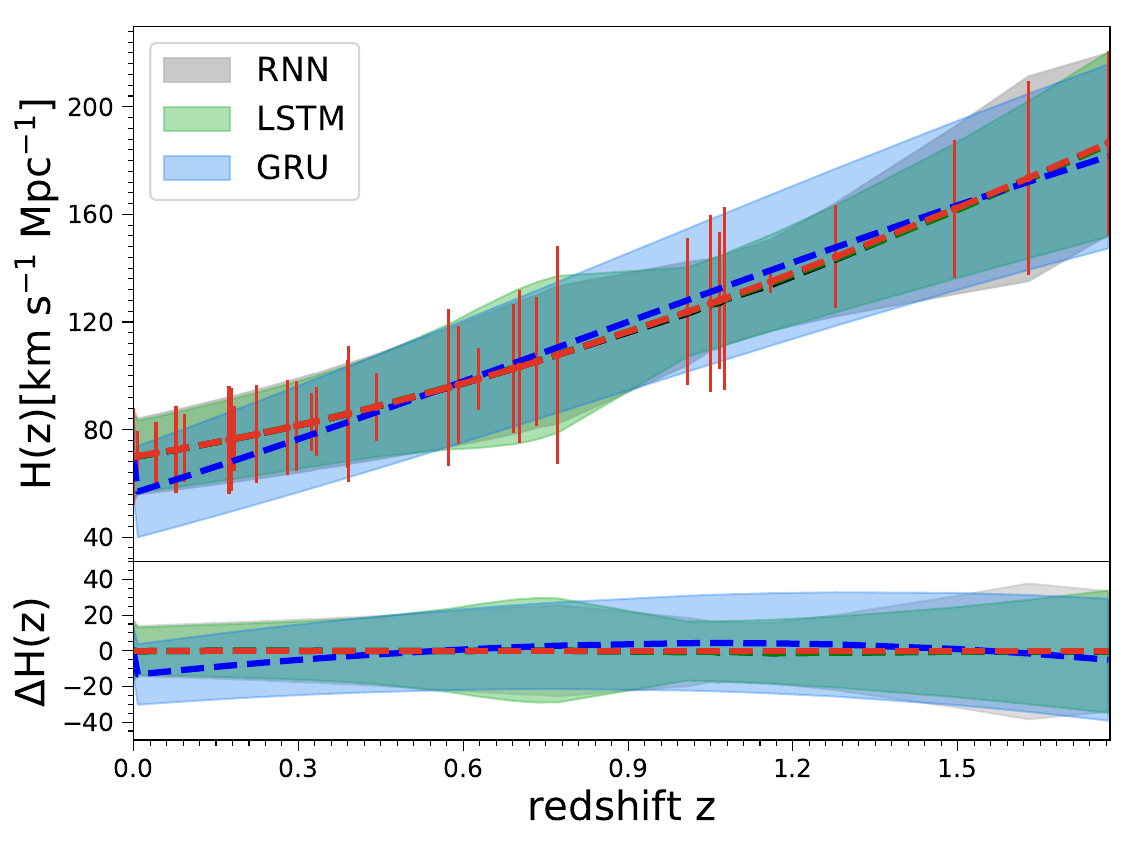}
	\caption{The same as Figure \ref{fig:sim_Hz_NoRandom_rec}, except now using the RNN, LSTM, and GRU network to reconstruct the Hubble parameter.}\label{fig:sim_Hz_NoRandom_rnn_lstm_gru}
\end{figure}
\begin{figure*}
	\centering
	\includegraphics[width=0.96\textwidth]{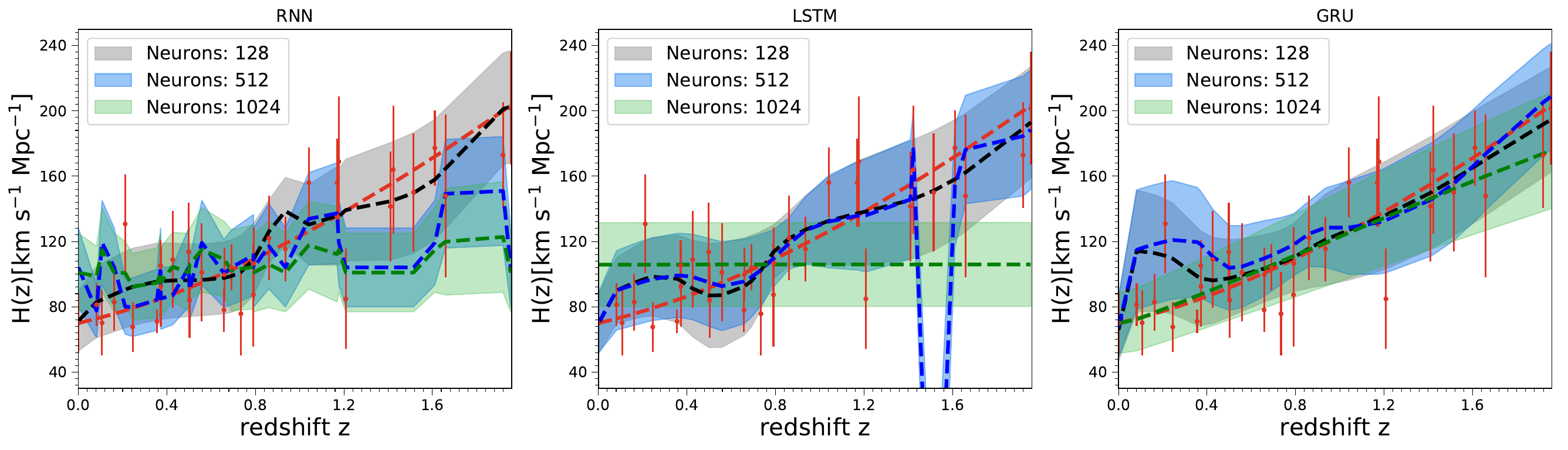}
	\caption{The same as Figure \ref{fig:mock_Hz_layer1}, except now using RNN, LSTM, and GRU to reconstruct the Hubble parameter.}\label{fig:mock_Hz_layer1_ann_rnn_bnn}
\end{figure*}

With the same procedure as in section \ref{sec:optimize_model}, we first find the optimal network models for RNN, LSTM, and GRU, respectively, by using the simulated Hubble parameter of Figure \ref{fig:mock_Hz}. The optimal models have one hidden layer and contain 128, 128, and 1024 neurons in the hidden layer for RNN, LSTM, and GRU, respectively. Then, we reconstruct functions of $H(z)$ from the simulated Hubble parameter of Figure \ref{fig:sim_Hz_NoRandom_rec}. The reconstructed functions of $H(z)$ with RNN, LSTM, and GRU are shown in Figure \ref{fig:sim_Hz_NoRandom_rnn_lstm_gru}. The black dashed line with shaded areas refer to the result of RNN, the green dashed line with shaded areas stand for the result of LSTM, and the blue dashed line with shaded areas are for that of GRU. We can see that the reconstructed functions for all three networks are consistent with the fiducial cosmological model (the red dashed line) within a $1\sigma$ confidence level. Moreover, the mean values of the reconstructed function are almost the same as the fiducial cosmological model for both RNN and LSTM, which is similar to that of the ANN method (see Figure \ref{fig:sim_Hz_NoRandom_rec}). These may indicate that RNN, LSTM, and GRU are capable of reconstructing functions from data.

However, the reconstructed functions are greatly influenced by the number of neurons in the hidden layer. In Figure \ref{fig:mock_Hz_layer1_ann_rnn_bnn}, we draw three reconstructed functions of $H(z)$ for RNN, LSTM, and GRU. These three functions of $H(z)$ are reconstructed with three network models that have different numbers of neurons in the hidden layer. For the RNN method (the left panel of Figure \ref{fig:mock_Hz_layer1_ann_rnn_bnn}), we can see that the reconstructed function of $H(z)$ will deviate from the fiducial cosmological model with the increase in the number of neurons in the hidden layer. This can also happen with the LSTM method (the middle panel of Figure \ref{fig:mock_Hz_layer1_ann_rnn_bnn}). Furthermore, when the number of neurons in the hidden layer is 1024, the reconstructed function of $H(z)$ using the LSTM method is slightly opposite in trend to the mock data, which is totally unreasonable. For the GRU method (the right panel of Figure \ref{fig:mock_Hz_layer1_ann_rnn_bnn}), the number of neurons in the hidden layer also affects the reconstructed functions of $H(z)$, even though the effect is slightly reduced.

The effect of the number of neurons in the hidden layer on the reconstructed functions of $H(z)$ indicates that it is not safe to reconstruct functions from the observational data with the optimal network model. Thus, this makes it difficult to reconstruct functions from observational data with RNN, LSTM, and GRU, respectively. Therefore, the ANN method is more reliable than RNN, LSTM, and GRU in the reconstruction of functions from observational data.

\section{Discussions}\label{sec:discussion}

\subsection{The ANN method}\label{sec:discussion_ANN}

In this work, the NN is designed to reconstruct functions of the Hubble parameter $H(z)$ and the luminosity distance $D_L(z)$ of SNe Ia. However, we note that it is mathematically proven that an NN with only one hidden layer can approximate any function with any accuracy if we use enough neurons \citep{Cybenko:1989,Hornik:1989}. Therefore, the ANN is a general method that can reconstruct functions for any kind of data.

There are many hidden parameters (or hyperparameters) in the NN, which should be selected before using ANN for the reconstruction of functions. In the process of supervised learning, the data are generally divided into three parts: the training set, the validation set, and the test set. The network models are trained on the training set, and the hidden parameters are optimized using the validation set. However, this training and evaluation strategy is not suitable in the task of reconstructing functions because all the data should be used to train the network to construct an approximate function, which means that the data cannot be divided to evaluate the network models. Thus, we present a new strategy to train and evaluate the network models in section \ref{sec:optimize_model}, by using simulated data.

In section \ref{sec:optimize_model}, we only consider optimizing the number of hidden layers and that of neurons in the hidden layer for the reconstruction of the Hubble parameter $H(z)$. The optimal network model selected in this work is applicable to both current and near-future observations of the Hubble parameter. However, we propose that the hidden parameters of the network should be optimized with the strategy in section \ref{sec:optimize_model} when the ANN method is used in other observational data sets. Furthermore, other hidden parameters of the network, such as batch normalization, learning rate, and batch size, can also be optimized using this strategy if one applies the ANN method to other similar tasks.

The ANN method proposed in this work can perform a reconstruction of a function from data without assuming a parameterization of the function, which is a completely data-driven approach. Moreover, the ANN method has no assumptions of Gaussian distribution for the random variables and can be used for any kind of data. We test the ANN method using both observational and simulated data in the sections \ref{sec:reconstruct_CCHz}, \ref{sec:reconstruct_DL} and \ref{sec:compare_other_NN}. The results indicate that the ANN method is reliable and unbiased, for both the best-fit values and errors of the reconstructed function. In addition, the reconstructed functions can be used to estimate cosmological parameters unbiasedly. Moreover, the results of section \ref{sec:reconstruct_CCHz} show that the ANN method is not sensitive to the input cosmology. Therefore, we propose that the ANN method will be a very promising method in the reconstruction of functions from data.

\subsection{Covariance matrix}\label{sec:discussion_cov}

The analysis in sections \ref{sec:reconstruct_CCHz} and \ref{sec:reconstruct_DL} shows that the function reconstructed by the ANN method can be used for further parameter estimation. Here, we illustrate a problem that should be noted when estimating parameters using the reconstructed function. For the given sample of observational Hubble parameters, we can train a network model to learn complex relationships between the redshift $z$ and $H(z)$ and its error. Then, any number of Hubble parameters can be obtained by feeding a sequence of redshifts to the network model. Obviously, this will lead to cosmological parameters being constrained to arbitrary precision when using many more samples than the observational data, which is unreasonable. Therefore, the number of samples generated by the reconstructed function should be similar to that of the observational data. We point this out to draw the attention of the reader if one uses the data generated from the reconstructed function in future research on parameter estimation.

\begin{figure}
	\centering
	\includegraphics[width=0.45\textwidth]{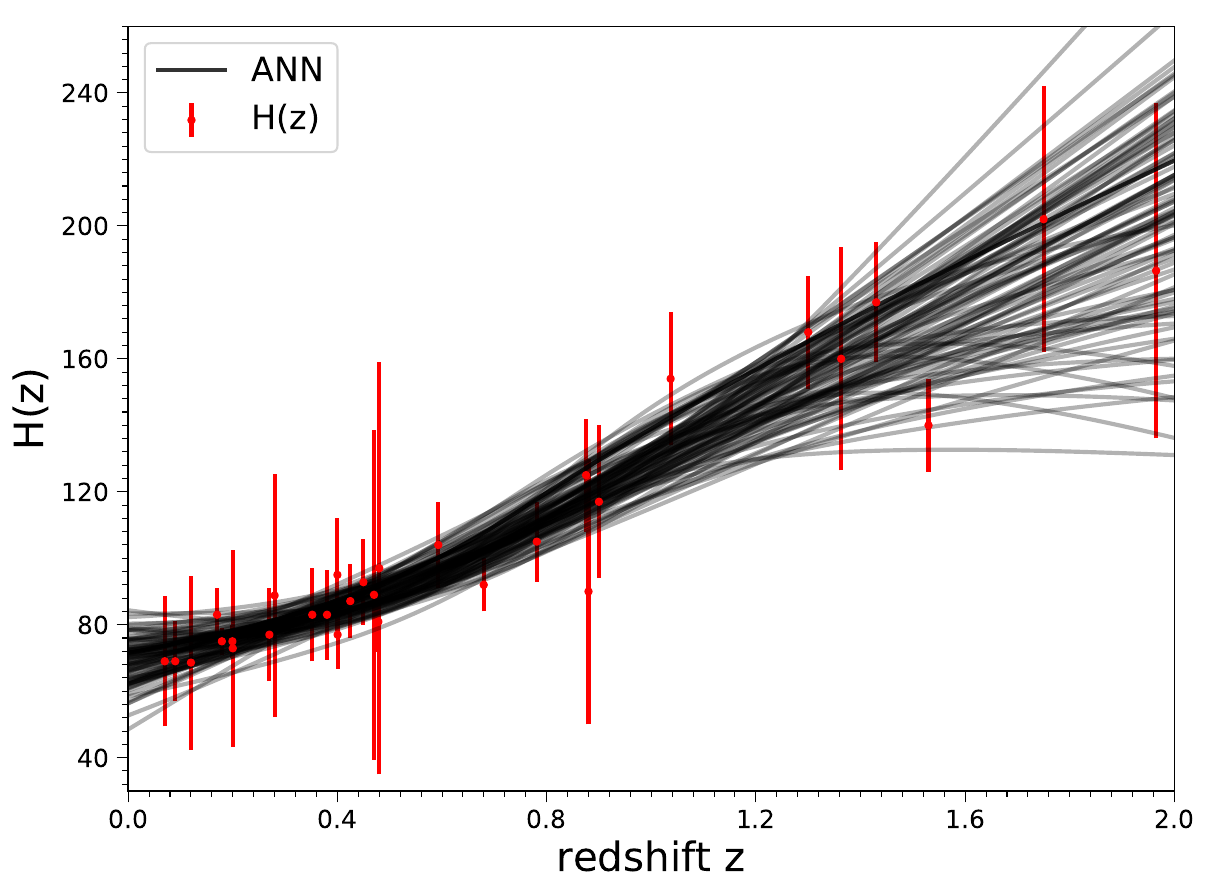}
	\caption{100 functions of $H(z)$ reconstructed by ANN.}\label{fig:multifunc_ann}
\end{figure}

The covariance matrix should be considered when using any number of samples in the parameter estimation. However, it is not easy to calculate the covariance matrix via the theory of ANNs. Therefore, here we provide a statistical method to calculate the covariance matrix. We take the Hubble parameter $H(z)$ as an example to illustrate the process of calculating the covariance matrix. Specifically, the key steps of this process are:
\begin{itemize}
	\item[1.] Generate 1000 realizations of a data-like sample by drawing $n$ $H(z)$ measurements via the Gaussian distribution $\mathcal{N}(H(z), \sigma^2_{H(z)})$, where $H(z)$ and $\sigma_{H(z)}$ are the observational Hubble parameter and corresponding errors, and $n$ is the number of data points in our observational data.
 	\item[2.] For each realization, train an ANN model using the corresponding samples of $H(z)$. Then, a function of $H(z)$ can be reconstructed using the trained ANN model. Note that errors of $H(z)$ are not reconstructed in this method, and thus there is only one neural in the output layer of the ANN model.
 	\item[3.] Obtain 1000 functions of $H(z)$ by repeating step 2. In Figure \ref{fig:multifunc_ann}, we show an example of 100 functions of $H(z)$.
 	\item[4.] For two Hubble parameters at the redshift $z_1$ and $z_2$, the covariance between them can be calculated by comparing the 1000 $H(z_1)$ and 1000 $H(z_2)$ values using
 	\begin{align}
 	\nonumber &{\rm\bf Cov}(H(z_i), H(z_j))\\
 	&=\frac{1}{N}\sum_{k=1}^N[(H(z_i)_k - \bar{H}(z_i))(H(z_j)_k - \bar{H}(z_j))],
 	\end{align}
 	where $N=1000$ is the number of $H(z)$ data points at the redshift $z_1$ or $z_2$, and $\bar{H}(z)$ is the average value of the 1000 $H(z)$ data points.
 	\item[5.] Using the method of step 4, the covariance between any two Hubble parameters with different redshifts can be calculated. Therefore, the covariance matrix can be further obtained. 	
\end{itemize}

\begin{figure}
	\centering
	\includegraphics[width=0.23\textwidth]{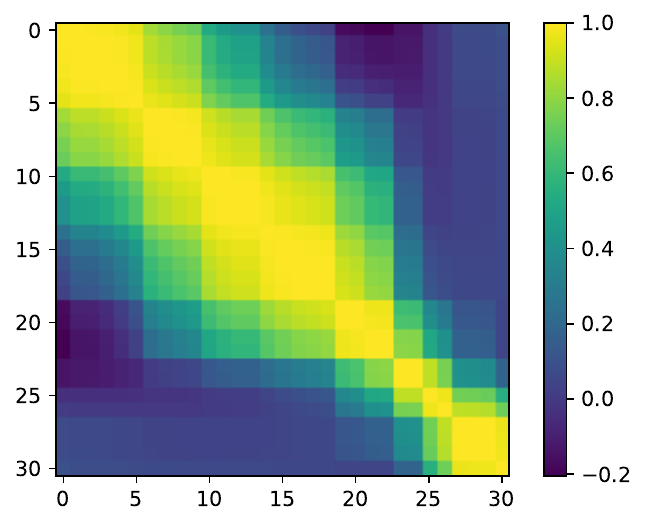}
	\includegraphics[width=0.23\textwidth]{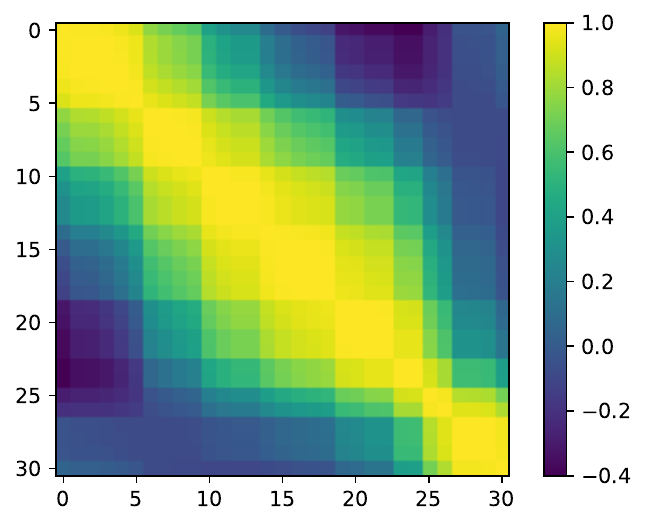}
	\caption{An example of the normalized covariance matrix of one set of samples of $H(z)$ reconstructed by ANN (left) and GP (right).}\label{fig:cov_ann_gp}
\end{figure}

Using this method, we reconstructed the Hubble parameter $H(z)$ and further obtained the corresponding covariance matrix. Here we normalize the covariance matrix to obtain the correlation coefficient
\begin{equation}
\rho = \frac{{\rm\bf Cov}(H(z_i), H(z_j))}{\sqrt{D(H(z_i))}\sqrt{D(H(z_j))}},
\end{equation}
where $D(H(z_i))$ is the variance of $H(z_i)$. We plot an example of the normalized covariance matrix in the left panel of Figure \ref{fig:cov_ann_gp}. At the same time, we also reconstruct the Hubble parameter using GP and the commonly used squared exponential covariance function is adopted:
\begin{equation}
k(x,\tilde{x}) =
\sigma_f^2 \exp\left( -\frac{(x - \tilde{x})^2}{2\ell^2} \right) \;,
\end{equation}
where $\sigma_f$ and $\ell$ are two hyperparameters that should be optimized. The corresponding normalized covariance matrix is further obtained, and an example is shown in the right panel of Figure \ref{fig:cov_ann_gp}. We can see that the covariance matrix obtained by ANN is similar to that of GP, which may indicate the rationality of the covariance matrix obtained by ANN. Therefore, this method may provide a possibility of calculating the covariance matrix for the function reconstructed by ANNs. However, the covariance matrix obtained in this way needs to be further researched in the parameter estimation, which will be shown in our future work.

\section{Conclusions}\label{sec:conclusion}

We propose that ANNs can be used to reconstruct functions from data. In this work, we test the ANN method using the Hubble parameter and SN Ia data by reconstructing functions of $H(z)$ and $D_L(z)$. We find that both $H(z)$ and $D_L(z)$ can be reconstructed with high accuracy, which indicates that the ANN method is a promising method in cosmological research. Furthermore, we also estimate parameters using the reconstructed functions of $H(z)$ and $D_L(z)$, and find the results are consistent with those obtained using the observational data directly. Therefore, we propose that the functions reconstructed by ANN can represent the actual distribution of observational data and can be used for parameter estimation in cosmological research. We will investigate these interesting issues in the future
works.

The ANN used in this work is a general method that could reconstruct a function from any kind of data without assuming a parameterization of the function, which is a completely data-driven approach. Moreover, this method has no assumptions of Gaussian distribution for the observational random variables; hence, it can be widely used in other observational data. Therefore, data-driven modeling based on the NN has the potential to play an important role in future cosmological research. Based on the ANN, a code for reconstructing functions from data was developed and can be downloaded freely.

\section{Acknowledgement} We thank the anonymous reviewer for his/her very enlightening comments. We thank Zhengxiang Li, Jingzhao Qi, Ping Guo, and Huan Zhou for helpful discussions. J.-Q. Xia is supported by the National Science Foundation of China under grants Nos. U1931202, 11633001, and 11690023, the National Key R\&D Program of China No. 2017YFA0402600, and the Fundamental Research Funds for the Central Universities, grant No. 2017EYT01.

\end{document}